\documentclass[12pt]{article}
\usepackage{amsmath,amssymb,exscale,cancel}
\usepackage{slashed}
\usepackage{graphicx}
\usepackage{epsfig}
\usepackage{multicol}
\usepackage{color}
\usepackage{mathrsfs}
\usepackage{blindtext}
 \usepackage{fancyhdr}
\usepackage{hyperref}
\usepackage{cite}
\usepackage{mathtools}

\usepackage{floatrow}

\usepackage{graphicx}
\usepackage{sidecap}

\usepackage{subfig}

\usepackage[latin1]{inputenc} 

\usepackage{multicol}  
 
 \usepackage{titlesec}
 
 \usepackage{rotating,slashed,xcolor,amsfonts,expdlist,charter}

\numberwithin{equation}{section}

\usepackage{xcolor}
\usepackage{sectsty}

\usepackage{mdframed}
\usepackage{titletoc}

\numberwithin{equation}{section}

\usepackage{xcolor}
\usepackage{sectsty}

\usepackage{mdframed}
\usepackage{titletoc}

\definecolor{secnum}{RGB}{13,151,225}
\definecolor{ptcbackground}{RGB}{212,237,252}
\definecolor{ptctitle}{RGB}{0,177,235}

\titlecontents{lsection}
  [5.8em]{\sffamily}
  {\color{secnum}\contentslabel{2.3em}\normalcolor}{}
  {\titlerule*[1000pc]{.}\contentspage\\\hspace*{-5.8em}\vspace*{5pt}%
    \color{white}\rule{\dimexpr\textwidth-15.5pt\relax}{1pt}}

\usepackage{hyperref}
\hypersetup{colorlinks,bookmarksopen,bookmarksnumbered,citecolor=rossos,
linkcolor=redy,pdfstartview=FitH,urlcolor=green-go}
\usepackage{slashed}

\definecolor{blus}{cmyk}{1,0.9,0,0.1}
\definecolor{verdes}{cmyk}{0.99,0,0.59,0.65}
\definecolor{rossos}{cmyk}{0,1,1,0.55}
\definecolor{redy}{cmyk}{0,1,1,0.7}
\definecolor{greeny}{cmyk}{0.99,0,0.59,0.98}
\definecolor{green-go}{cmyk}{0.79,0,0.59,0.5}

\usepackage{titlesec}

\newcommand{\beq}{\begin{equation}}
\newcommand{\eeq}{\end{equation}}

\def\hhref#1{\href{http://arxiv.org/abs/#1}{arXiv:#1}} 

\newcommand{\tmtextbf}[1]{{\bfseries{#1}}}
\newcommand{\tmtextrm}[1]{{\rmfamily{#1}}}
\newcommand{\bp}{M_P}

\def\be{\begin{equation}}
\def\ee{\end{equation}}
\def\ba{\begin{array} }
\newcommand{\Tr}{\,{\rm Tr}}
\def\bac{\begin{array} {c}}
\def\bacc{\begin{array} {cc}}
\def\baccc{\begin{array} {ccc}}
\def\bacccc{\begin{array} {cccc}}
\def\ea{\end{array}}
\def\bea{\begin{eqnarray}}
\def\eea{\end{eqnarray}}

\definecolor{red}{rgb}{1,0,0}

\def\psl{\hbox{\hbox{${p}$}}\kern-1.9mm{\hbox{${/}$}}}
\def\dsl{\hbox{\hbox{${\partial}$}}\kern-2.2mm{\hbox{${/}$}}}
\def\Dsl{\hbox{\hbox{${D}$}}\kern-2.6mm{\hbox{${/}$}}}

\newcommand{\gappeq}{{\rlap{{\raise}.5ex\text{\ensuremath{>}}}{{\lower}.5ex\text{\ensuremath{\sim}}}}}
\newcommand{\lappeq}{{\rlap{{\raise}.5ex\text{\ensuremath{<}}}{{\lower}.5ex\text{\ensuremath{\sim}}}}}
\newcommand{\I}{\tmtextrm{1{\kern}-.24em l}}

\begin{document}
\topmargin -1.0cm
\oddsidemargin 0.9cm
\evensidemargin -0.5cm

{\vspace{-1cm}}
\begin{center}

\vspace{-1cm}

 {\Huge \tmtextbf{ 
\color{rossos} \hspace{-0.9cm}  
Natural Metric-Affine Inflation \hspace{-1.6cm}}} {\vspace{.5cm}}\\

\vspace{1.3cm}

{\large {\bf Antonio Racioppi}$^a$ and {\bf Alberto Salvio}$^{b,c}$
{\em  
\vspace{.4cm}

${}^a$ National Institute of Chemical Physics and Biophysics, R\"avala 10, 10143 Tallinn, Estonia

\vspace{0.6cm}

${}^b$ Physics Department, University of Rome Tor Vergata, \\ 
via della Ricerca Scientifica, I-00133 Rome, Italy\\

\vspace{0.6cm}

${}^c$ I. N. F. N. -  Rome Tor Vergata,\\
via della Ricerca Scientifica, I-00133 Rome, Italy\\

\vspace{.4cm}


\vspace{0.4cm}

\vspace{0.2cm}

 \vspace{0.5cm}
}

\vspace{.cm}

}
\vspace{0.7cm}

\end{center}

\noindent ---------------------------------------------------------------------------------------------------------------------------------
\begin{center}
{\bf \large Abstract}
\end{center}
\noindent We consider here natural inflation in the low energy (two-derivative) metric-affine theory containing only the minimal degrees of freedom in the inflationary sector, i.e.~the massless graviton and the pseudo-Nambu-Goldstone boson (PNGB). This theory contains the Ricci-like and parity-odd Holst  invariants together with non-minimal couplings between the PNGB and the above-mentioned invariants.  The Palatini and Einstein-Cartan realizations of natural inflation are particular cases of our construction. Explicit models of this type featuring non-minimal couplings are shown to emerge from the microscopic dynamics of a QCD-like theory with an either sub-Planckian or trans-Planckian confining scale and that is renormalizable on Minkowski spacetime. Moreover, for these models,  we find regions of the parameter space where the inflationary predictions agree with the most recent observations at the $2\sigma$ level. We find that in order to enter the $1\sigma$ region it is necessary (and sufficient) to have a finite value of the Barbero-Immirzi parameter and a sizable non-minimal coupling between the inflaton and the Holst invariant (with sign opposite to the Barbero-Immirzi parameter). Indeed, in this case the potential of the canonically normalized inflaton develops a plateau as shown analytically.

\vspace{0.5cm}

\noindent---------------------------------------------------------------------------------------------------------------------------------

%

  \vspace{-0.9cm}
  
  \newpage 
\tableofcontents

\noindent

\vspace{0.5cm}

\section{Introduction}\label{intro}

 Inflation is currently accepted as the best theory of the earliest observable stages of our universe. It explains the horizon and flatness problems and provides us with a microscopic origin for the small inhomogeneities and anisotropies in the cosmic microwave background (CMB). Such fluctuations include tensor perturbations that may show up in the future as primordial gravitational waves in detectors like the DECi-hertz Interferometer Gravitational wave Observatory (DECIGO)~\cite{DECIGO}, the Big Bang Observer (BBO)~\cite{Crowder:2005nr,BBO2}  and the Advanced Laser Interferometer Antenna (ALIA)~\cite{Gong:2014mca,Crowder:2005nr}.  In practice, inflation is often realized by introducing a scalar field, the inflaton, with a potential that is nearly flat for a large enough range of field values. Indeed, this acts approximately as a vacuum energy and the universe inflates. 

 From the particle-theory point of view, however, it is difficult to realize such potential flatness without introducing a high level of fine tuning. One way to efficiently circumvent this problem is to impose a symmetry that protects the inflaton potential from large radiative corrections. Natural inflation~\cite{Freese:1990rb} is a concrete way to realize this idea: the inflaton is identified with the PNGB of an approximate and spontaneously broken symmetry. In the limit where the explicit symmetry-breaking terms vanish the PNGB becomes an exact Nambu-Goldstone boson and, therefore, acquires an exact shift symmetry, which is the manifestation of Goldstone's theorem. As a result the flatness of the inflaton potential turns out to be natural in 't Hooft's sense~\cite{tHooft:1979rat}. 
 
 Another field-theoretic attractive feature of natural inflation is the fact that it admits an explicit UV completion within an asymptotically free theory. One considers a QCD-like theory with a large (Planckian) confining scale and identifies the inflaton with the composite particle analogous to the lightest meson\footnote{See also~\cite{Salvio:2019wcp} for an analysis of reheating in this scenario.}~\cite{Freese:1990rb,Salvio:2021lka}. The effective inflaton action should then be periodic in the inflaton field\footnote{See, however, Refs.~\cite{Bostan:2022swq,AlHallak:2022haa} for some studies of non-periodic effective actions, whose UV completion is difficult to conceive.}. So far a simple cosine potential and non-minimal coupling between the Ricci scalar and the inflaton has been shown to emerge from such a UV complete theory~\cite{Salvio:2021lka}.  This has been achieved by assuming gravity to be formulated ab initio in the ``metric" way: namely, the (gravitational) connection was assumed to be equal to the Levi-Civita (LC) one, $\Gamma_{\mu~\sigma}^{~\,\rho}$. The corresponding inflationary predictions were obtained in~\cite{Salvio:2023cry}.
 
 It is, however, well known that, in a purely geometric formulation of gravity, the metric, $g_{\mu\nu}$, and the connection, ${\cal A}_{\mu~\sigma}^{~\,\rho}$, can be independent and so generically ${\cal A}_{\mu~\sigma}^{~\,\rho}\neq\Gamma_{\mu~\sigma}^{~\,\rho}$. Such formulation, known as metric-affine gravity, can be used to find physically different actions for the metric and the matter fields.  For example, metric-affine gravity admits two rather than just one two-derivative curvature invariants: the usual Ricci-like scalar and the Holst invariant~\cite{Hojman:1980kv,Nelson:1980ph,Holst:1995pc}, which can be used to construct new models\footnote{For works in the so-called Palatini formulation, which is a particular case of metric-affine gravity as we recall in Sec.~\ref{The action}, see e.g.~\cite{Jarv:2017azx,Jarv:2020qqm,Gialamas:2023flv,Barker:2024ydb} and refs.~therein. }~\cite{Hecht:1996np,BeltranJimenez:2019hrm,Pradisi:2022nmh,Salvio:2022suk,DiMarco:2023ncs}. In particular, this scenario may then be used to find more UV complete realizations of natural inflation.
 
 Thus, the purpose of this paper is to study natural inflation in metric-affine gravity focusing on models that admit a UV completion.
This alternative realizations of natural inflation is also motivated by the fact that a simple minimally-coupled (PNGB) inflaton with a cosine potential is in tension with  the most recent CMB observations performed by the Planck, Background Imaging of Cosmic Extragalactic Polarization (BICEP) and Keck collaborations~\cite{Ade:2015lrj,Planck2018:inflation,BICEP:2021xfz}, within Einstein's general relativity.  Other possible ways of restoring the agreement between natural inflation and CMB observations have been discussed in~\cite{Achucarro:2015rfa,Ferreira:2018nav,Antoniadis:2018yfq,Salvio:2019wcp,Salvio:2021lka,Simeon:2020lkd,Salvio:2022mld,Salvio:2023cry}. 

The paper is organized as follows. In Sec.~\ref{The action} we construct the low-energy metric-affine theory for the metric and the PNGB inflaton. For the sake of minimality, we restrict the analysis to metric-affine models where the field content taking part in inflation includes only the graviton and a single PNGB. In Appendix~\ref{micro} we also investigate possible UV completions of this effective action. Sec.~\ref{sec:potential} is devoted to an analytical study of the corresponding effective potential of the canonically normalized inflaton. The inflationary predictions are then analyzed in Sec.~\ref{INFp}. Finally, in Sec.~\ref{Conclusions} we offer a detailed summary of the results and our conclusions.

\section{The action and the dynamical degrees of freedom}\label{The action}

 Here we build a natural inflationary scenario in a metric-affine theory containing only the minimal degrees of freedom in the inflationary sector (the massless graviton and the PNGB). 
 So we demand the distorsion, i.e.~the difference between the full connection and the Levi-Civita one
 \be C_{\mu~\sigma}^{~\,\rho} \equiv {\cal A}_{\mu~\sigma}^{~\,\rho}-\Gamma_{\mu~\sigma}^{~\,\rho}, \ee
 to be non dynamical.  The  torsion $T_{\mu\nu\rho}$ is defined in terms of the distorsion by $T_{\mu\nu\rho} \equiv C_{\mu\nu\rho}-C_{\rho\nu\mu},$. 
 
 In this case, as shown in~Ref.~\cite{Pradisi:2022nmh}, the general effective action including the metric, the distorsion and all matter fields (scalars $\phi$, fermions $\psi$ and gauge fields $A^I_\mu$ with field strength $F^I_{\mu\nu}$) in the theory is  
 \be S_{\rm EFT} = \int d^4x\sqrt{-g}\left({\cal F}_{\mu\nu\rho\sigma} {\cal T}^{\mu\nu\rho\sigma}(\Phi) + \Sigma(\Phi, {\cal D}\Phi, C)    \right). \label{Seq1}\ee
Here ${\cal F}_{\mu\nu~~\sigma}^{~~~\rho}$ is the curvature associated with\footnote{As usual, we use $g_{\mu\nu}$ and its inverse $g^{\mu\nu}$ to lower and raise the spacetime indices.} ${\cal A}_{\mu~\sigma}^{~\,\rho}$,
\be {\cal F}_{\mu\nu~~\sigma}^{~~~\rho} \equiv \partial_\mu{\cal A}_{\nu~\sigma}^{~\,\rho}-\partial_\nu{\cal A}_{\mu~\sigma}^{~\,\rho}+{\cal A}_{\mu~\lambda}^{~\,\rho}{\cal A}_{\nu~\sigma}^{~\,\lambda}-{\cal A}_{\nu~\lambda}^{~\,\rho}{\cal A}_{\mu~\sigma}^{~\,\lambda},\ee
$\Phi$ represents the set of fields that are independent of $C_{\mu~\sigma}^{~\,\rho}$, namely
\be \Phi=\{g_{\mu\nu}, \phi, \psi, F_{\mu\nu}^I, ... \},\ee
the dots are curvatures and covariant derivatives of the previous fields constructed with the LC connection, 
${\cal T}^{\mu\nu\rho\sigma}(\Phi)$ is a rank-four contravariant tensor that depends on $\Phi$ only (not on its derivatives) and  $\Sigma(\Phi, {\cal D}\Phi, C)$ is a scalar that depends on $\Phi$ 
and  $C_{\mu~\sigma}^{~\,\rho}$ only. In the dependence of $\Sigma$ we have stressed that the covariant derivative ${\cal D}\Phi$ built with ${\cal A}_{\mu~\sigma}^{~\,\rho}$ can appear, but this can be expressed in terms of $C$ and $\Phi$. Note that ${\cal T}^{\mu\nu\rho\sigma}(\Phi)$ and $\Sigma(\Phi, {\cal D}\Phi, C)$ should also be invariant under all gauge and global symmetries present in the theory. 

Einstein-Cartan and Palatini gravitational theories are particular cases of the more general metric-affine construction, when $\{{\cal D}_\mu g_{\nu\rho} =0,T_{\mu\nu\rho}\neq 0\}$ 
and $\{{\cal D}_\mu g_{\nu\rho} \neq 0,T_{\mu\nu\rho}= 0\}$, respectively.

Let us now consider the part of $S_{\rm EFT}$ that contains only the massless graviton and the PNGB. Restricting ourselves to terms that feature only two derivatives, which are expected to be the leading ones in the low energy limit, we obtain the following action of natural metric-affine inflation: 
\be S_{\rm NI}= \int d^4x\sqrt{-g}\left[\alpha(\phi){\cal R}+\beta(\phi)\tilde{\cal R}   -\frac{\partial_\mu \phi \, \partial^\mu \phi}{2} - V(\phi) \right], \label{SNIstart} \ee
where here $\phi$ is the PNGB field,
\begin{equation}
    V(\phi) \equiv \Lambda^4 \left[ 1 + \cos\left(\frac{\phi}{f}\right) \right]+\Lambda_\mathrm{cc}
    \label{eq:V(phi)}
\end{equation}
is the natural-inflaton potential~\cite{Freese:1990rb}, $\Lambda$ and $f$ are two energy scales and  $\Lambda_{\rm cc}$ accounts for 
the (tiny and positive) cosmological constant responsible for the observed dark energy and is negligible during inflation, which occurs at a much larger energy scale. In~(\ref{SNIstart}) $\alpha$ and $\beta$ are (not yet specified) functions of $\phi$ and ${\cal R}$ and $\tilde{\cal R}$ are, respectively, a scalar and pseudoscalar  contraction of the curvature,
\be {\cal R} \equiv {\cal F}_{\mu\nu}^{~~~\mu\nu}, \qquad  \tilde{\cal R} \equiv \frac1{\sqrt{-g}}\epsilon^{\mu\nu\rho\sigma}{\cal F}_{\mu\nu\rho\sigma},\label{RRpdef}\ee
where $\epsilon^{\mu\nu\rho\sigma}$ is the totally antisymmetric Levi-Civita symbol with $\epsilon^{0123}=1$. We refer to $\tilde{\cal R}$ as the Holst invariant (see~\cite{Hojman:1980kv,Nelson:1980ph,Holst:1995pc}). The pseudoscalar $\tilde{\cal R}$ vanishes for $C_{\mu~\sigma}^{~\,\rho}=0$ (that is when the connection is the LC one) thanks to the cyclicity property $R_{\mu\nu\rho\sigma}+R_{\nu\sigma\rho\mu}+R_{\sigma\mu\rho\nu}=0$, where $R_{\mu\nu\rho\sigma}$ represents the Riemann tensor.
 This is the reason why in standard Riemannian geometry $\tilde{\cal R}$ is absent. Therefore, $\tilde{\cal R}$ can be considered as a manifestation of a connection that is independent of the metric.

The distorsion is non dynamical (does not introduce additional degrees of freedom) if and only if the action has the form in~(\ref{Seq1})~\cite{Pradisi:2022nmh}.  So, in our case the distorsion can be integrated out to obtain an equivalent metric theory, i.e.~a gravity theory where the connection is the LC one and the components of the distorsion are expressed in terms of the other fields, $\Phi$. Performing this integration for the inflationary action in~(\ref{SNIstart}) leads to~\cite{Karananas:2021zkl,BeltranJimenez:2019hrm,Pradisi:2022nmh} 
\be S_{\rm NI}= \int d^4x\sqrt{-g}\left[\alpha R   -\frac{\partial_\mu \phi \, \partial^\mu \phi}{2} +\frac{\frac{\alpha}4\partial_\mu\alpha\partial^\mu\alpha-\alpha\partial_\mu\beta\partial^\mu\beta 
+ 2\beta\partial_\mu\alpha\partial^\mu\beta}{\frac23\left(\beta^2+\frac{\alpha^2}4\right) } - V\right], \ee
where $R$ is the Ricci scalar 
(remember that $R=\cal R$ when $ {\cal A}_{\mu~\sigma}^{~\,\rho} = \Gamma_{\mu~\sigma}^{~\,\rho} $).

In the expression above we have a non-minimal coupling $\alpha(\phi)R$. This can be removed by performing the Weyl transformation 
\be g_{\mu\nu}\to \frac{\bp^2}{2\alpha(\phi)}g_{\mu\nu}, 
\ee
where $\bp$ is the reduced Planck mass. 
Such transformation requires $\alpha>0$ in order for the transformation to be regular and to preserve the signature of the metric.
One then obtains
\bea S_{\rm NI} &=& \int d^4x\sqrt{-g}\left[\frac{\bp^2}{2} R -\frac{\bp^2}{4\alpha}\partial_\mu \phi \, \partial^\mu \phi -\frac{3\bp^2}{4\alpha^2}\partial_\mu\alpha\partial^\mu\alpha  \right. \nonumber\\ &&\left. + \frac{\frac{\partial_\mu\alpha\partial^\mu\alpha}8-\frac{\partial_\mu\beta\partial^\mu\beta}2
+\frac{\beta}{\alpha}\partial_\mu\alpha\partial^\mu\beta}{\frac2{3\bp^2}\left(\beta^2+\frac{\alpha^2}4\right) } - \frac{\bp^4V}{4\alpha^2} \right]. \eea
This action can be rewritten as follows
\be S_{\rm NI} =\int d^4x\sqrt{-g}\left[\frac{\bp^2}{2} R -\frac{k(\phi)}{2}\partial_\mu \phi \, \partial^\mu \phi - \frac{V(\phi)}{F^2(\phi)} \right],\label{SNIE}\ee 
where $F(\phi)\equiv 2\alpha(\phi)/M_P^2$ and 
\bea
    k(\phi) &\equiv& \frac{M_P^2}{2}\left[\frac1\alpha+\frac{3\alpha'^2}{\alpha^2}\left(1-\frac{\alpha^2}{\alpha^2+4\beta^2}\right) 
    +\frac{12\beta'^2}{\alpha^2+4\beta^2}
    -\frac{24\beta\alpha'\beta'}{\alpha(\alpha^2+4\beta^2)}\right] \\ &=& \frac{M_P^2}{2}\left[\frac1\alpha+\frac{12(\alpha' \beta-\alpha\beta')^2}{\alpha^2(\alpha^2+4\beta^2)}\right] > 0,
    \label{eq:K(phi)}
\eea
where a prime represents a derivative with respect to $\phi$ (from now on a prime on a function represents a derivative with respect to its argument). Since $k(\phi)$ is always positive, $\phi$ is never a ghost.

Note that in the particular case $\beta = 0$, the expression above simply becomes
\begin{equation}
    k(\phi) = \frac{1}{F(\phi)}\equiv \frac{M_P^2}{2\alpha(\phi)}.
    \label{eq:K(phi)b0}
\end{equation}
In the case of natural inflation formulated directly in the metric formalism (as opposed to the metric-affine formalism), namely starting from the action
\be \int d^4x\sqrt{-g}\left[\alpha(\phi) R -\frac1{2}\partial_\mu \phi \, \partial^\mu \phi - V(\phi) \right]\label{metricNI}\ee 
and then going to the Einstein frame, one finds~(\ref{SNIE}) but with~\cite{Salvio:2023cry}
\begin{equation}
    k(\phi) \equiv \frac{2F(\phi)+3M_P^2F'^2(\phi)}{2F^2(\phi)}= \frac{1}{F(\phi)}+\frac{3M_P^2F'^2(\phi)}{2F^2(\phi)}.
    \label{eq:K(phi)2}
\end{equation}
 This $k(\phi)$ differs from the one in~(\ref{eq:K(phi)b0}) because of the extra positive term $3M_P^2F'^2(\phi)/(2F^2(\phi))$. Therefore, we can see that the two theories differ even if one sets $\beta=0$ in the metric-affine case.

The scalar kinetic term in~(\ref{SNIE}) can be brought into a canonical form through a field redefinition $\phi\to \chi$ satisfying 
\begin{equation}
    \frac{d\chi}{d\phi} \,\equiv\, \sqrt{k(\phi)}, \qquad \chi(\phi=0) = 0.
    \label{eq:define.Chi}
\end{equation}
Since $k(\phi)$ is always positive, $\chi$ is a monotonically growing function of $\phi$ and can be inverted to obtain $\phi$ as a function of $\chi$. Then we can express the action in terms of the metric and $\chi$:
\begin{equation}
    S_{\rm NI} = \int d^4x\sqrt{-g}\, \left\{\,
    \frac{1}{2}M_P^2R \,-\, 
    \frac{1}{2}
   g^{\mu\nu}\partial_\mu\chi\partial_\nu\chi
    \,-\,
    U(\chi) \right\}, \label{Eaction}
\end{equation}
where we have defined  
\begin{equation}
    U(\chi) \,\equiv\, \frac{V(\phi(\chi))}{F^2(\phi(\chi))}.
    \label{eq:U(chi)}
\end{equation}

The inflationary predictions of the theory can be now extracted from the action in the form~(\ref{Eaction}) through standard methods once the functions $\alpha(\phi)$ and $\beta(\phi)$ are specified.

In Appendix~\ref{micro} we provide a microscopic origin for the following $\alpha$ and $\beta$,
\be \alpha(\phi) = \frac{\bp^2}{2}\left[1+\xi \left(1+\cos\left(\frac{\phi}{f}\right)\right)\right], \qquad \beta(\phi) = \beta_0+\frac{\bp^2}2 \tilde\xi \left(\cos\left(\frac{\phi}{f}\right)+1\right)\label{alphabetaUV}\ee 
This choice is, therefore, well motivated from the particle-physics perspective and we will thus assume it in the rest of the paper. The quantity $\bp^2/(4\beta_0)$ is called the  Barbero-Immirzi parameter~\cite{Immirzi:1996di,Immirzi:1996dr}.

\section{An analytical study of the inflationary potential} \label{sec:potential}

As we have seen in the previous section, the inflationary potential $U(\chi)$ (for the canonically normalized inflaton $\chi$), defined in~(\ref{eq:U(chi)}) is the result of an integration and an inversion of a function. In our context we are unable to perform these operations analytically, expect in some very specific cases. Nevertheless, we provide an analytical discussion here in order to obtain an understanding of the numerical calculations that will be reported in Sec.~\ref{INFp}.

First of all, we start spending a few words on $U$ as a function of $\phi$ and then we move to discuss the impact of the canonical normalization of the scalar field. For generic $\xi$, but neglecting the tiny $\Lambda_{\rm cc}$, \eqref{eq:U(chi)}  becomes
\begin{equation}
  U(\chi) = \frac{\Lambda ^4 \left(1+\cos \left(\frac{\phi(\chi) }{f}\right)\right)}{\left[ 1+ \xi \left(1+ \cos \left(\frac{\phi(\chi)
   }{f}\right) \right) \right]^2}.
  \label{eq:U:phi}
\end{equation}
\begin{figure}[t]
    \centering
    \includegraphics[width=0.5\textwidth]{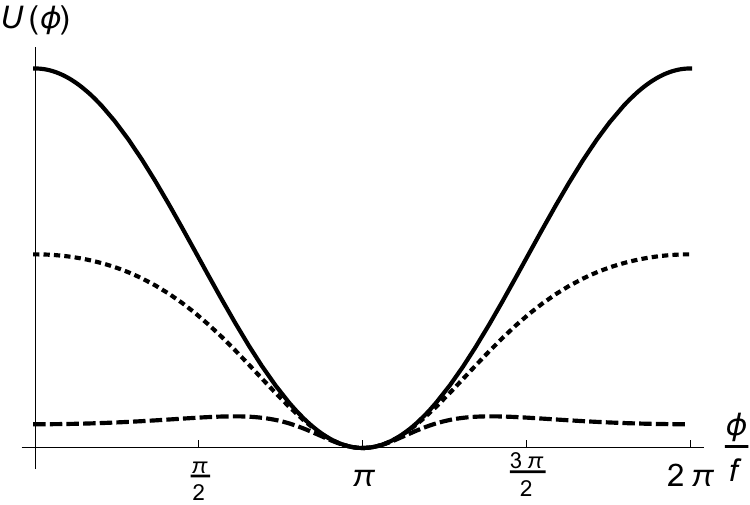}
    \caption{\em Reference plots for $U(\phi)$ in the domain $0 \leq \frac{\phi}{f} \leq 2\pi$ for $\xi=0$ (continuous), $0 < \xi < \frac{1}{2}$ (dotted) and $\xi>\frac{1}{2}$ (dashed).}
    \label{fig:U:phi}
\end{figure}

Some reference plots are given in Fig.~\ref{fig:U:phi}. 
Since the potential is  an even function of $\phi$ with period $2\pi f$, let us focus on the $\phi \in [0,\pi f]$ domain. 

The quantity in~(\ref{eq:U:phi}) has been studied as function of $\phi$ in~\cite{Salvio:2023cry}. The stationary points are
\begin{eqnarray}
\phi_1 &=& 0 \\
\phi_2 &=& \pi  f\end{eqnarray}
and, when $\xi>1/2$,
\be 
\phi_3 = f \arccos \left(\frac{1-\xi}{\xi} \right).\ee
The value of the second derivative on those points is
\begin{eqnarray}
U_{\phi\phi}(\phi_1) &=& \frac{\Lambda ^4 (2 \xi -1)}{f^2 (2 \xi +1)^3}, \\
U_{\phi\phi}(\phi_2) &=& \frac{\Lambda ^4}{f^2},\\
U_{\phi\phi}(\phi_3) &=& \frac{\Lambda ^4 (1-2 \xi )}{8 f^2 \xi }.
\end{eqnarray}
So, $\phi_2$  is always a minimum. The stationary point $\phi_3$  is always a maximum.  For $\xi > \frac{1}{2}$ the value $\phi_1=0$ is a minimum, otherwise it is a maximum. In the latter case  inflation happens as usual from the maximum to the minimum. We stress that for $\xi=\frac{1}{2}$, $\phi_1=\phi_3=0$ is a maximum. On the other hand, when $\xi > \frac{1}{2}$, there seems to be two possible regions available for inflation: $0<\phi<\phi_3$ and $\phi_3 < \phi < \phi_2$. In the latter inflation happens as before just with the replacing of 0 with $\phi_3$. In the former, one could think to inflate from the maximum in $\phi_3$ to the minimum in $\phi_1=0$ and then use $U(\phi_1)$ as the cosmological constant. Unfortunately, the potential value in those points
\begin{eqnarray}
U(\phi_1) &=& \frac{2 \Lambda ^4}{(2 \xi +1)^2}, \label{eq:U:max} \\
U(\phi_3) &=& \frac{\Lambda ^4}{4 \xi }
\end{eqnarray}
is so that there is only a factor of $\xi$ (in the big $\xi$ limit) of difference between the two scales i.e.~not enough to account for the inflationary scale and the cosmological constant at the same time. Therefore, from now on we will consider only inflation happening towards the minimum in $\phi_2$. More details about $U$ will be given in the next section.

Let us now proceed with the discussion of the effect of the change of field from $\phi$ to $\chi$.
As discussed around~(\ref{eq:define.Chi}) the kinetic function, $k(\phi)$, defines the canonically normalized inflaton, $\chi(\phi)$, and the inverse function $\phi(\chi)$. The latter, as well as the potential $V(\phi)/F^2(\phi)$, determines the potential of the canonically normalized field, $U(\chi)$, through~(\ref{eq:U(chi)}). Therefore, $k(\phi)$ is a key ingredient to determine the inflationary observables and in this section we study it for the theoretically well-motivated values of $\alpha(\phi)$ and $\beta(\phi)$ in~(\ref{alphabetaUV}).

\begin{figure}[t!]
\begin{center}
\hspace{-0.37cm}
  \includegraphics[scale=0.5]{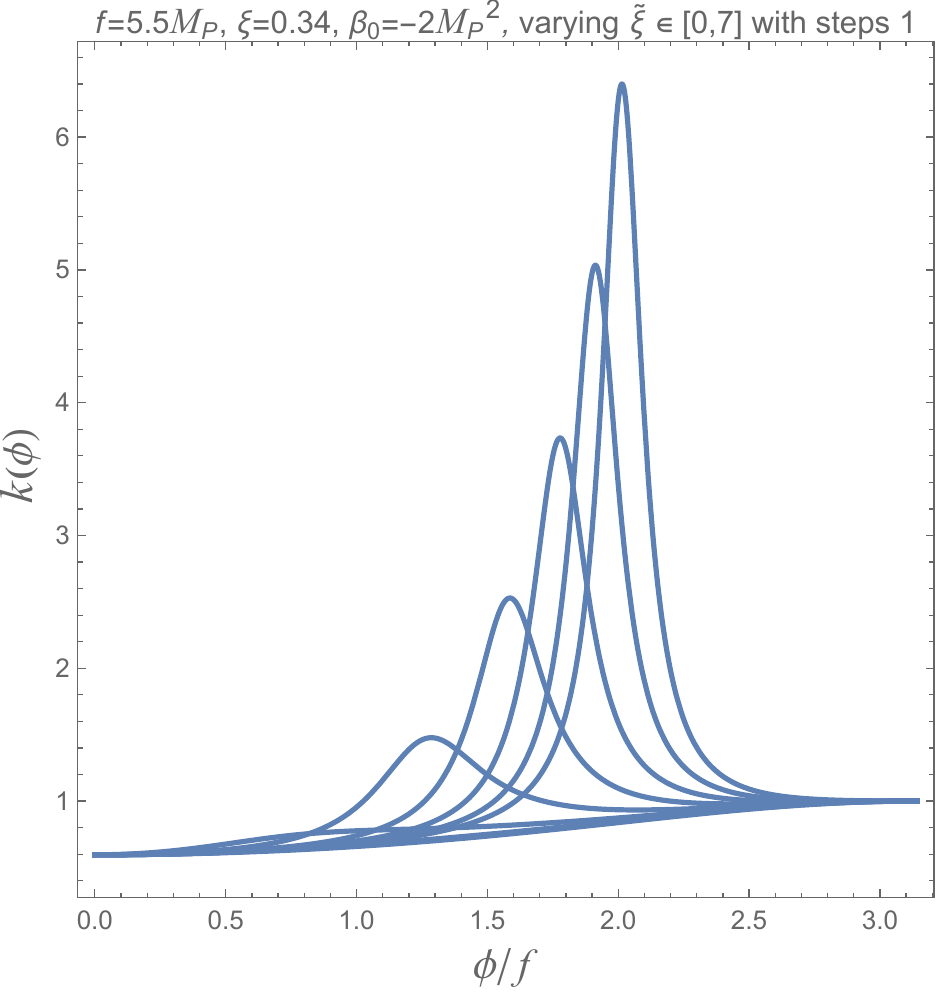} \qquad
 \includegraphics[scale=0.5]{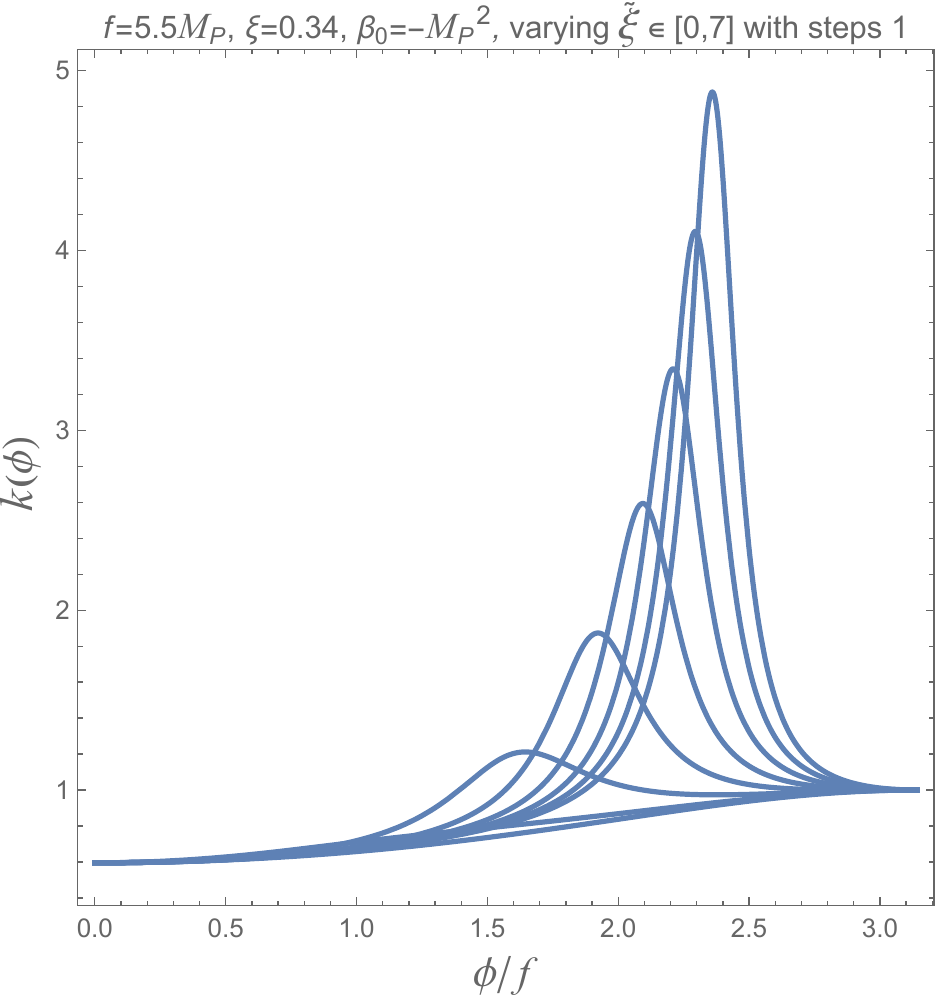}\\
\hspace{-0.4cm} \includegraphics[scale=0.52]{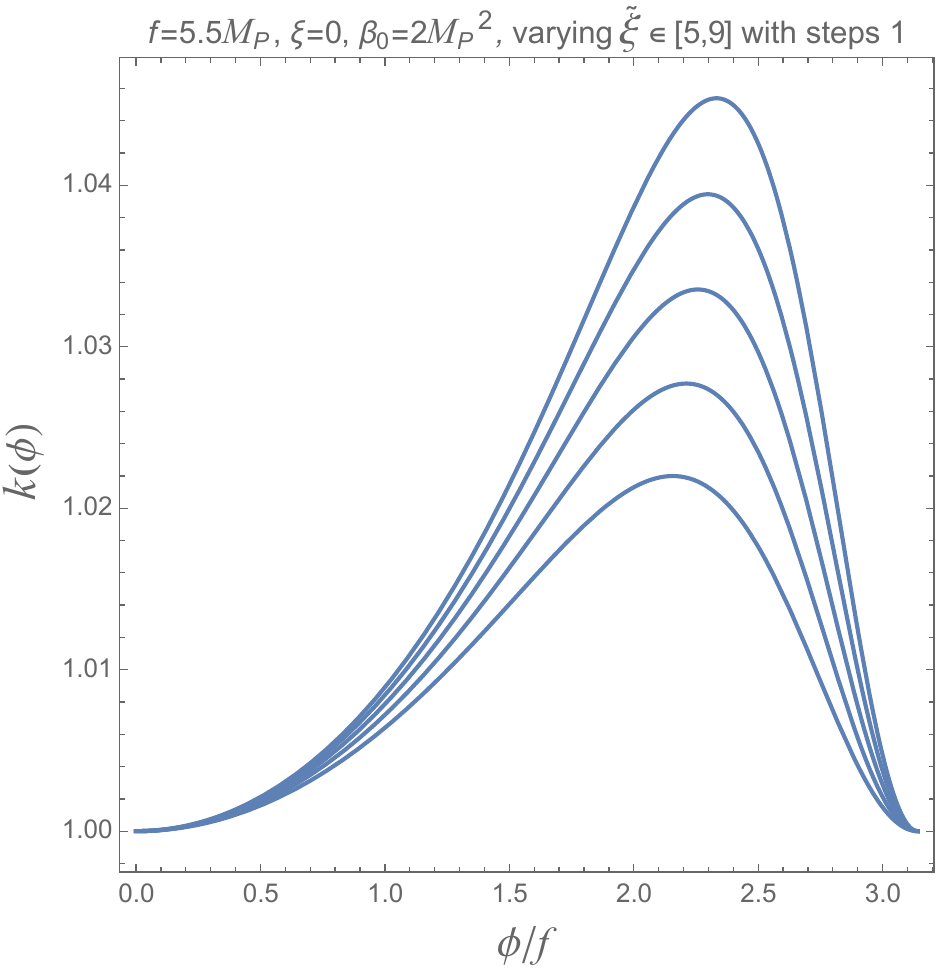} \qquad
 \includegraphics[scale=0.52]{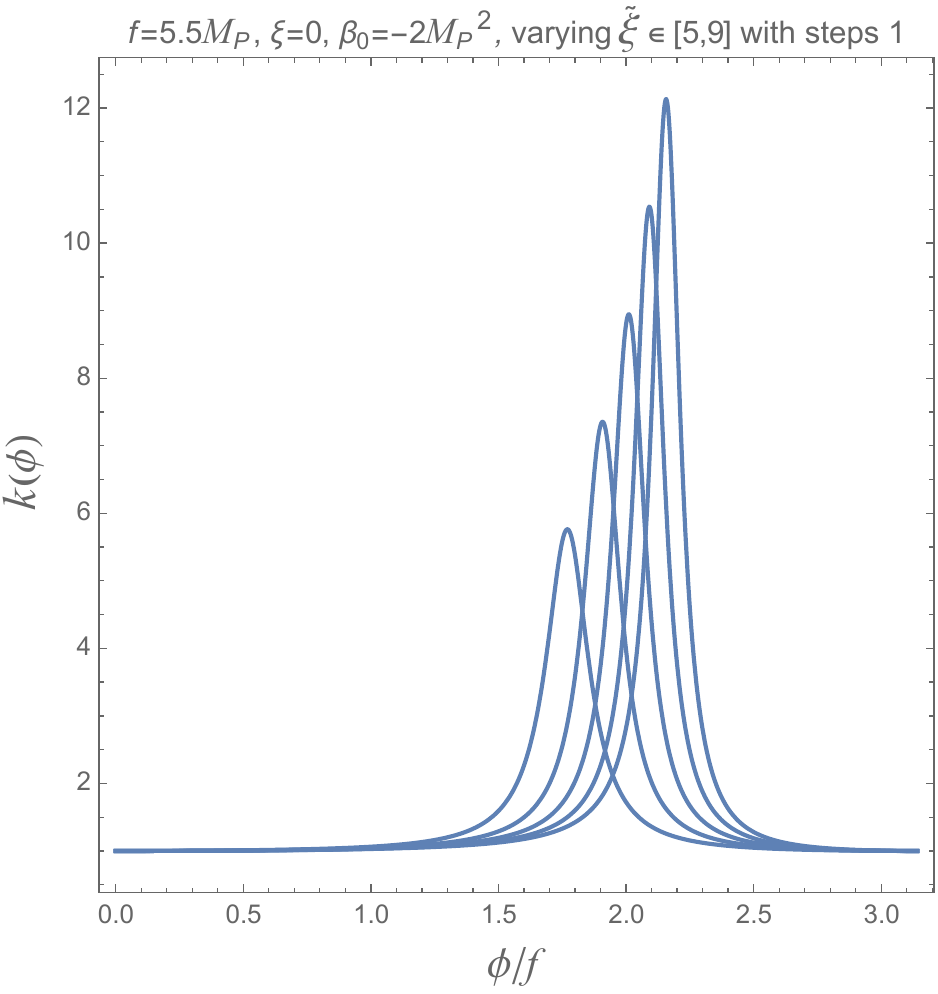} 
 \end{center} 
   \caption{\em  The kinetic function $k(\phi)$. A peak appears increasing $\tilde\xi$.}
\label{kofphi}
\end{figure}

 In Fig.~\ref{kofphi} the function $k(\phi)$ defined in Eq.~(\ref{eq:K(phi)}) is plotted. We find that a peak appears in $k$ increasing $\tilde\xi$ (the higher $\tilde\xi$ the higher the peak). For $\phi$ around the location of the peak, $\phi_{\rm peak}$, the function $\chi(\phi)$ increases more rapidly. This corresponds to a quasi-flat region  in the function $\phi(\chi)$: those values of $\chi$ that are mapped to a $\phi$ around $\phi_{\rm peak}$. 
Eq.~(\ref{eq:U(chi)}) tells us that a quasi-flat region in the function $\phi(\chi)$ is also a quasi-flat region in the potential of the canonically normalized inflaton, $U(\chi)$. The higher the peak in $k(\phi)$ is the flatter and larger this quasi-flat region turns out to be. This analytical understanding will be checked through numerical calculations in Sec.~\ref{INFp}, and the plateau in $U(\chi)$ will help in reaching the agreement with CMB observations. It is, therefore, interesting to analytically study how the location and height of the peak depends on the parameters of the model. 

For the sake of simplicity we explicitly present this analysis in the case $\xi=0$, for which $\alpha = M_P^2/2$ and thus $F=1$. For any $\beta(\phi)$, Eq.~(\ref{eq:K(phi)}) then tells us that the kinetic function is 
\be k(\phi) = 1+\frac{6 M_P^2 \beta '(\phi )^2}{\frac{M_P^4}{4}+4 \beta (\phi )^2}. \label{kxi0}\ee
Setting $\beta(\phi)$ as in~(\ref{alphabetaUV}), we can now look for the location of $\phi_{\rm peak}$ by requiring that $k'(\phi_{\rm peak})=0$.
Barring values of $\phi$ such that $\beta '(\phi )=0$, which are points of minimum according to~(\ref{kxi0}), the equation $k'(\phi)=0$ leads to
a second-order algebraic equation for $\cos(\phi/f)$. There is only one admissible solution (respecting  $|\cos(\phi/f)|\leq 1$), which gives
\be \cos\left(\frac{\phi_{\rm peak}}{f}\right) 
= \frac{\sqrt{\cal B}\sqrt{{\cal B}+2{\cal C}} - \cal B-\cal C}{\cal C}, 
\qquad  ({\cal B} \equiv M_P^4 + 16\beta_0^2, 
\quad {\cal C} \equiv 8M_P^2\tilde\xi(M_P^2 \tilde\xi + 2 \beta_0)). \label{eq:cosphipeak}
\ee 
Note that 
\be {\cal B}+2{\cal C} \geq 16\beta_0^2+16M_P^2\tilde\xi(M_P^2 \tilde\xi + 2 \beta_0)\geq 16 (\beta_0+M_P^2\tilde\xi)^2 \geq0, \label{Bp2C}\ee 
therefore, the square root in~(\ref{eq:cosphipeak}) is always real. From~(\ref{Bp2C}) one can also easily show that the modulus of the right-hand side of~(\ref{eq:cosphipeak}) is always less than or equal to one and thus can be equal to a cosine.

We can now determine the height of the peak by inserting this result in~(\ref{kxi0}):
\be k(\phi_{\rm peak}) = 1+\frac{3 M_P^6 \tilde\xi^2}{2 f^2} \, \,  \frac{1- \cos^2\left(\frac{\phi_{\rm peak}}{f}\right)}{\frac{M_P^4}{4}+ 4 \beta (\phi_{\rm peak})^2}.\label{kphipeak}\ee 

Note that there is a symmetry $\beta(\phi) \to -\beta(\phi)$ (see Eqs.~\eqref{eq:K(phi)} and~\eqref{kxi0}). Therefore, only the relative sign between $\beta_0$ and $\tilde\xi$ matters. From now on we will use the convention where $\beta_0$ can change sign, while $\tilde\xi$ remains positive.

\begin{figure}[t!]
\begin{center}
 \includegraphics[scale=0.487]{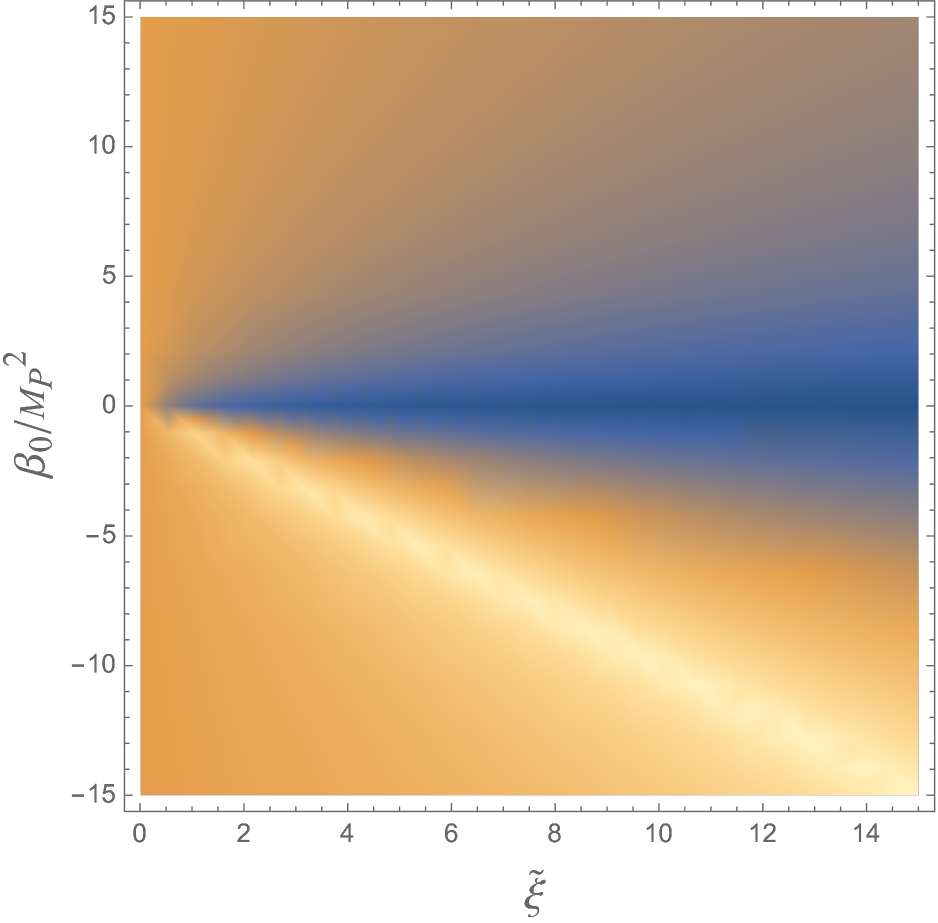}
 \includegraphics[scale=0.487]{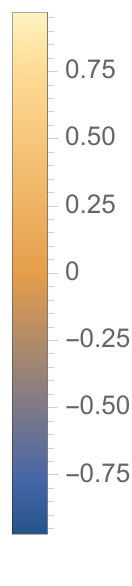}
 \includegraphics[scale=0.487]{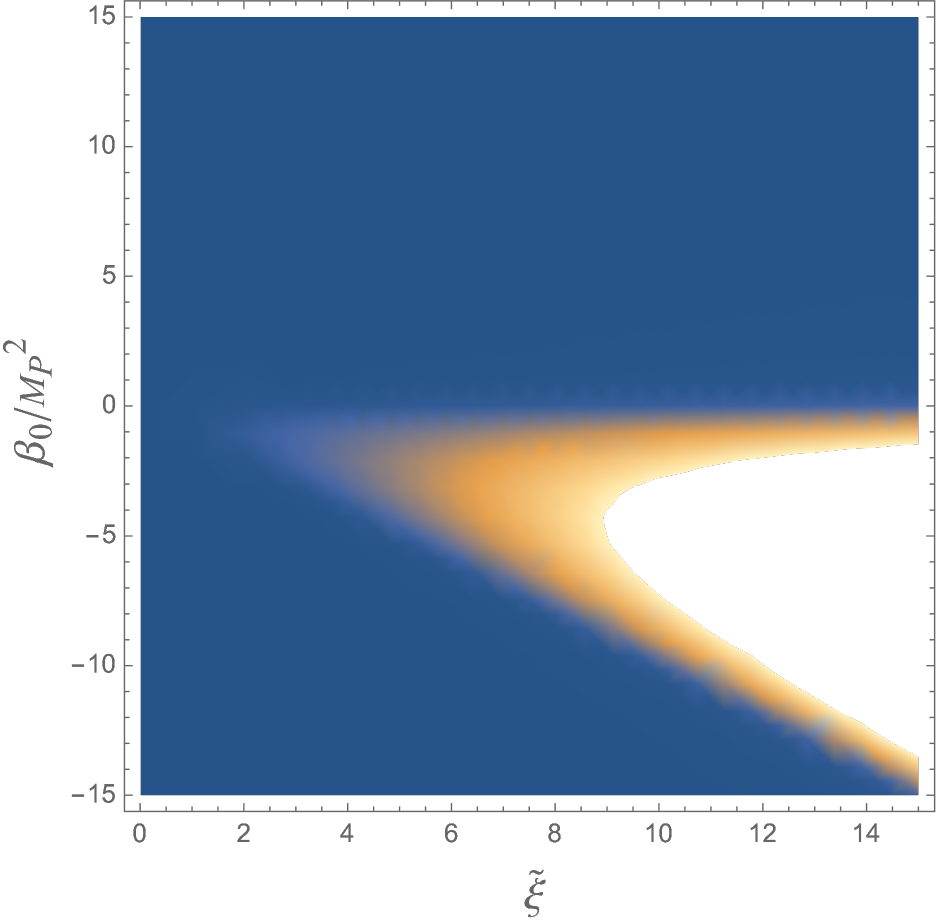} 
 \includegraphics[scale=0.487]{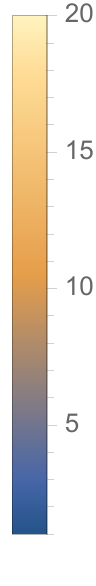}
 \end{center} 
   \caption{\em {\bf Left plot:} the location of the peak of the kinetic function, $\cos(\phi_{\rm peak}/f)$, for $\xi = 0$. {\bf Right plot:} the height of the peak, $k(\phi_{\rm peak})$, for $\xi = 0$ and $f = 5M_P$.}
\label{peaklochight}
\end{figure}

In Fig.~\ref{peaklochight} we show $\cos(\phi_{\rm peak}/f)$ and $k(\phi_{\rm peak})$ as functions of $\tilde\xi$ and $\beta_0$. The quantity $\cos(\phi_{\rm peak}/f)$ is independent of $f$, while $k(\phi_{\rm peak})$ does depend on it and we set $f=5M_P$ in Fig.~\ref{peaklochight}; however, as clear from Eq.~(\ref{kphipeak}), for a high peak, $k(\phi_{\rm peak})$ simply scales as $1/f^2$.

Before ending this section let us observe that in Fig.~\ref{kofphi} we have chosen trans-Planckian values of $f$. As we will see in Sec.~\ref{INFp}, among others, it is also for such values that we reach the agreement with CMB observations. Nevertheless, it is important to note that this setup do not necessarily invalidate the effective-field-theory treatment of gravity, which we are relying on. Indeed, the condition that allows us to do so is that the energy density should be much less than the Planck energy density \cite{WeinbergC}
\be |U(\chi)| \ll \bp^4, \label{eftC}\ee
for all relevant values of $\chi$. As we will show in Sec.~\ref{INFp}, this condition is amply satisfied. It should be kept in mind, however, than in some UV completions of all interactions such as string theory~\cite{Banks:2003sx} having a trans-Planckian $f$ can spoil the validity of natural inflation, at least in its simplest incarnation~\cite{Kim:2004rp,Dimopoulos:2005ac} (see~\cite{Baumann:2014nda} for a review). In some others, such as  asymptotically free/safe theories~\cite{WeinbergAS} 
 featuring higher-derivative terms 
 (see e.g.~\cite{Salvio:2024joi} and references therein) a trans-Planckian $f$ can be consistent~\cite{Salvio:2019wcp}. We find this aspect interesting as observing the predictions of natural inflation may give us information on the theory that UV completes Einstein gravity. For this reason, in Sec.~\ref{INFp} we will consider both trans-Planckian and sub-Planckian values of $f$ and, as we will show, agreement with data can be achieved in both cases. 

\section{Inflationary predictions} \label{INFp}
In this section we present the inflationary predictions of the model. Using the slow-roll approximation, all the inflationary observables can be derived from the inflaton Einstein-frame potential \eqref{eq:U(chi)} and its derivatives.  
The potential slow-roll parameters are defined as:
\bea
\epsilon_U  (\chi) &=& \frac{M_P^2}{2}\left(\frac{U'(\chi)}{U(\chi)}\right)^2 \, , \label{eq:epsilon}
\\
\eta_U  (\chi) &=& M_P^2 \frac{U''(\chi)}{U(\chi)} \, , 
\eea
\bea
\xi^2_U (\chi) &=& M_P^4 \frac{U'(\chi) U'''(\chi)}{U(\chi)^2} \, .
\eea
The expansion of the Universe is evaluated in number of e-folds, which is 
\be
N_e =  \frac{1}{M_P^2} \int_{\chi_{\textrm{end}}}^{\chi_N} {\rm d}\chi \, \frac{U(\chi)}{U'(\chi)} ,
\label{eq:Ne}
\ee
where the field value at the end of inflation is given by $\epsilon  (\chi_{\textrm{end}}) = 1 $, while the field value $\chi_N$ at the time a given scale left the horizon is given by the corresponding $N_e$. 
The tensor-to-scalar ratio $r$, the spectral index $n_\textrm{s}$ and its running $\alpha_\text{s} \equiv \text{d} n_\text{s}/ \text{d} \ln k$ are:
\bea
r  &=& 16\epsilon_U  (\chi_N) \,  , \label{eq:r} \\
n_\textrm{s}  &=& 1+2\eta_U  (\chi_N)-6\epsilon_U  (\chi_N) \, ,  \label{eq:ns} \\
\alpha_\textrm{s}  &=& 16 \epsilon_U (\chi_N) \eta_U (\chi_N) - 24 \epsilon_U^2 (\chi_N) - 2 \xi^2_U (\chi_N) \, . \label{eq:alphas}
\eea
Finally, the amplitude of the scalar power spectrum is
\be
 A _\textrm{s} = \frac{1}{24 \pi^2 M_P^4}\frac{U(\chi_N)}{\epsilon_U  (\chi_N)} \, ,
 \label{eq:As:th}
\ee
which needs to satisfy the experimental constraint~\cite{Planck2018:inflation}
\be
 A _\textrm{s} \simeq 2.1 \times 10^{-9} \, .
 \label{eq:As:exp}
\ee
In order to get a better understanding of the impact of $\alpha(\phi)$ and $\beta(\phi)$ we consider three different scenarios:
\begin{itemize}
 \item[1.]  $\xi \neq 0$ and $\beta(\phi)=0$.
 \item[2.]  $\xi=0$ and $\tilde\xi >0$. Note that we cannot set $\alpha(\phi)=0$ (see for instance Eqs.~\eqref{SNIstart} and~\eqref{eq:K(phi)}).
 \item[3.]  $\xi>0$ and $\tilde\xi >0$.
\end{itemize}

\subsection{$\xi \neq 0$ and $\beta(\phi)=0$}
This is the scenario that reproduces the non-minimal natural inflation without the Holst invariant. Before starting our analysis we introduce the following two parameters
\be
 \delta_\Lambda \equiv \frac{\Lambda}{M_P} \, , \qquad \delta_f \equiv \frac{f}{M_P} \, ,
\label{eq:deltas}
\ee
that allow to measure the natural inflation mass scales $\Lambda$ and $f$ in terms of $M_P$.

\subsubsection{$\xi < 0$ and $\beta(\phi)=0$}

\begin{figure}[t!]
     \subfloat[]{\includegraphics[width=0.45\textwidth]{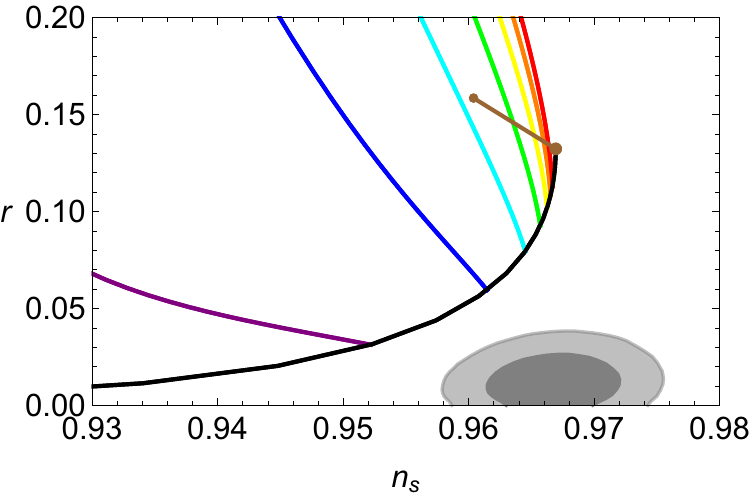}}%
    \subfloat[]{\includegraphics[width=0.45\textwidth]{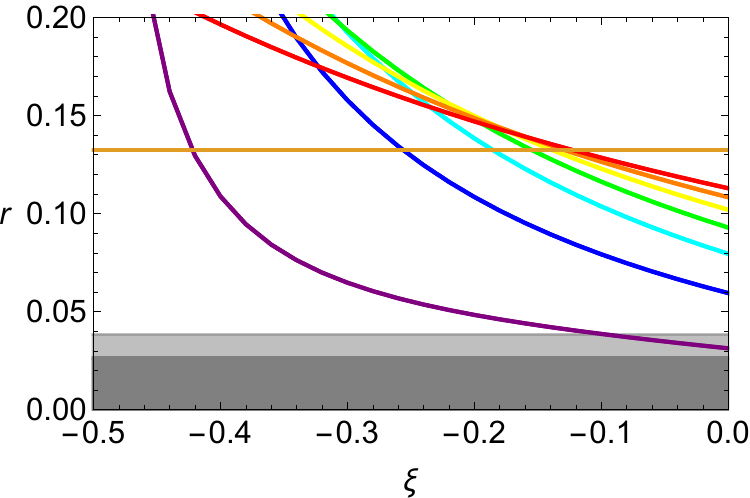}}%

    \subfloat[]{\includegraphics[width=0.45\textwidth]{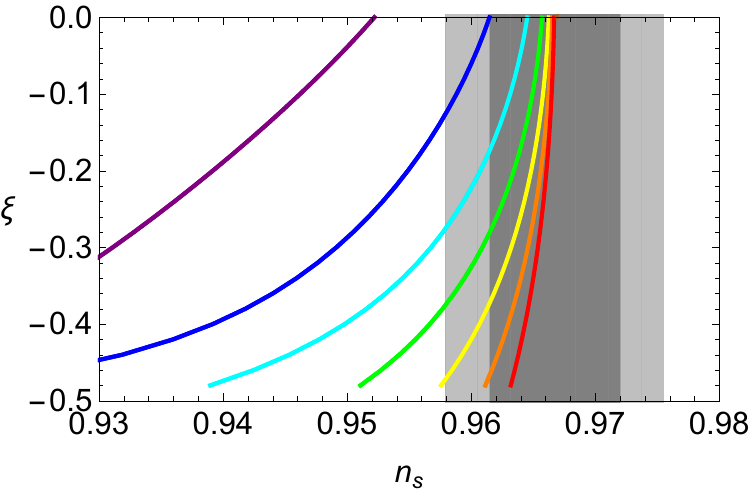}}%
    \subfloat[]{\includegraphics[width=0.45\textwidth]{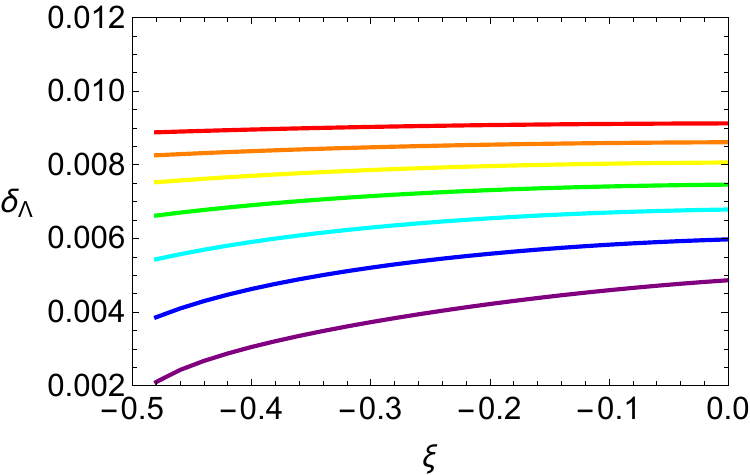}}%

    \caption{\em  $r$ vs.~$n_\text{s}$ (a),  $r$ vs.~$\xi$ (b),  $\xi$ vs.~$n_\text{s}$ (c),  $\delta_\Lambda$ vs.~$\xi$ (d) for $N_e=60$ and $\delta_f$ ranging from 5 (purple) to 14 (red) with steps of 1.5, displayed in rainbow colors. The gray areas represent the 1,2$\sigma$ allowed regions coming  from  the latest combination of Planck, BICEP/Keck and baryon-acoustic-oscillation (BAO) data~\cite{BICEP:2021xfz}. For reference the predictions of quadratic inflation for $N_e \in [50,60]$ (brown) and standard natural inflation for $N_e=60$ (black).} 
    \label{fig:results:Palatini:minus}%
\end{figure}

As long as $\alpha(\phi)$ in Eq.~\eqref{alphabetaUV} stays strictly positive, $\xi$ is allowed to take negative values. This sets the lower bound $\xi > -\frac{1}{2}$.

In Fig.~\ref{fig:results:Palatini:minus} we plot the corresponding results for $r$ and $n_\text{s}$ versus the parameters $\xi$ and $\delta_f$ when $N_e=60$  and $\delta_\Lambda$ is fixed so that the constraint \eqref{eq:As:exp} is satisfied.


We can see that with $\xi$ increasing in absolute value $r$ ($n_\text{s}$) increases (decreases) leading the prediction even more away from the allowed region. This was somehow expected because a negative $\xi$ increases the height of maximum of $U$ in Eq.~\eqref{eq:U:max} and this usually comes with an increase in $r$. Therefore, we conclude that the configuration with $\xi < 0$ and $\beta(\phi)=0$ is excluded by data 
and the computation of the predictions for $\alpha_\text{s}$ is not needed.

\begin{figure}[p!]
     \subfloat[]{\includegraphics[width=0.45\textwidth]{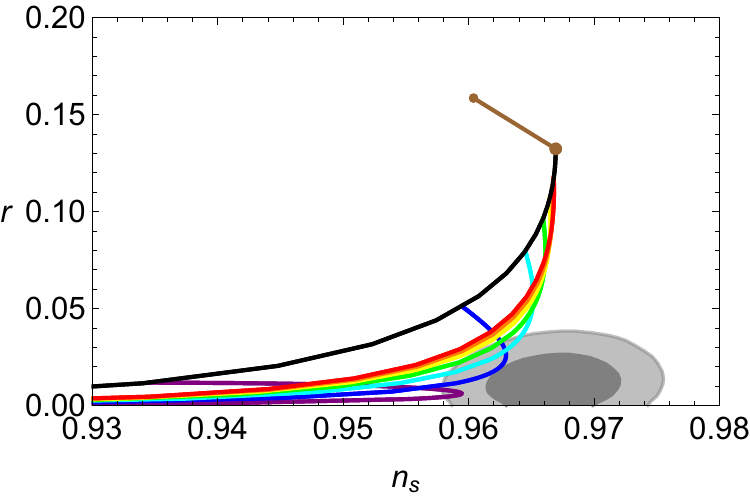}}%
    \subfloat[]{\includegraphics[width=0.45\textwidth]{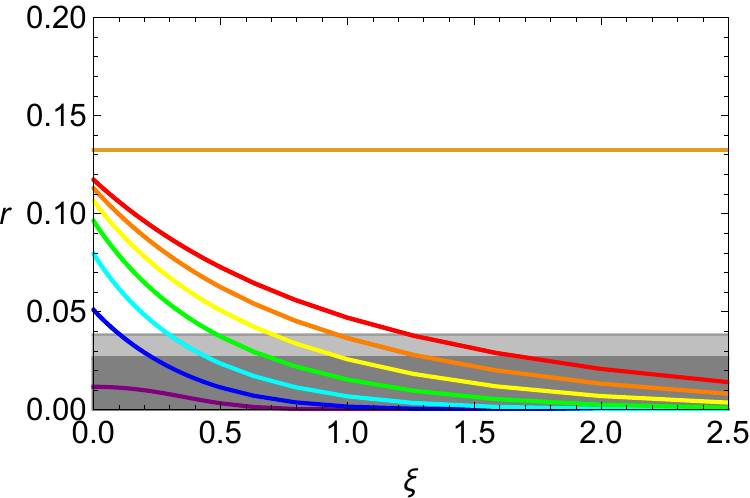}}%

    \hspace*{0.1cm} \subfloat[]{\includegraphics[width=0.44\textwidth]{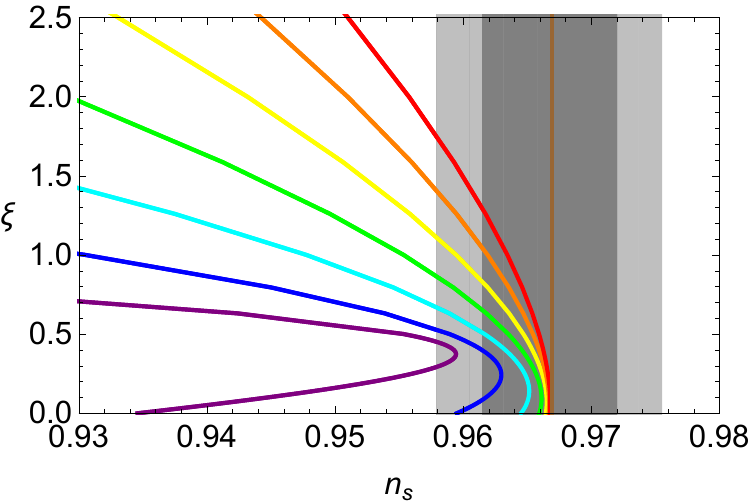}}%
    \subfloat[]{\includegraphics[width=0.45\textwidth]{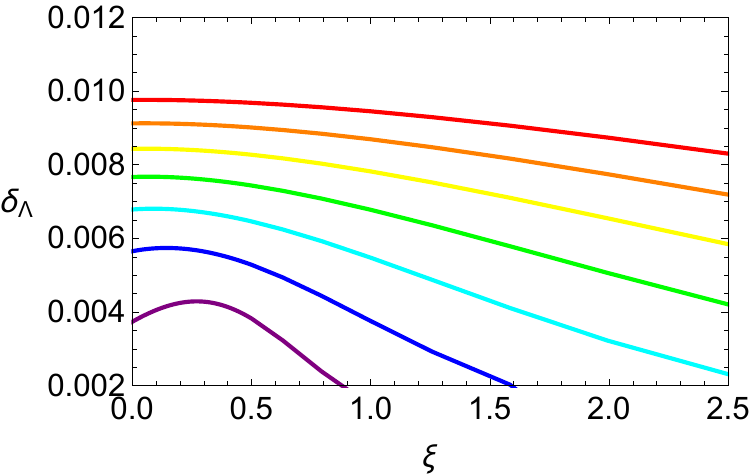}}%

    \caption{\em $r$ vs.~$n_\text{s}$ (a),  $r$ vs.~$\xi$ (b),  $\xi$ vs.~$n_\text{s}$ (c),  $\delta_\Lambda$ vs.~$\xi$ (d) for $N_e=60$ and $\delta_f$ ranging from 4 (purple) to 16 (red) with steps of 2, displayed in rainbow colors. The gray, brown and black color codes are the same as in Fig.~\ref{fig:results:Palatini:minus}.}
    %
    \label{fig:results:Palatini}%
\end{figure}
\begin{figure}[p!]
     \subfloat[]{\includegraphics[width=0.5\textwidth]{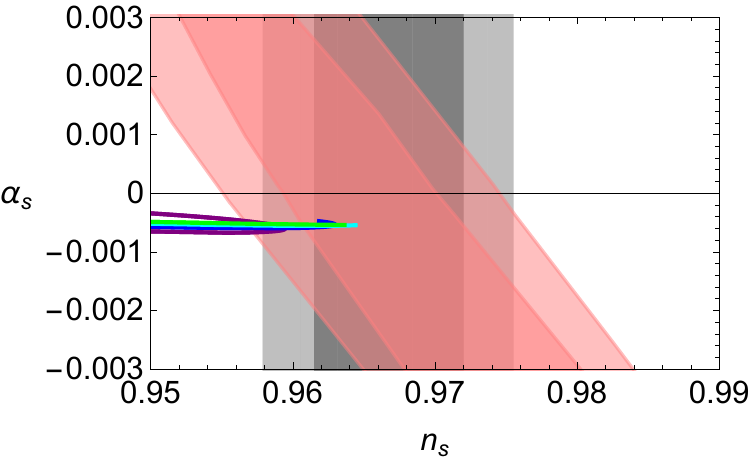}}%
    \subfloat[]{\includegraphics[width=0.5\textwidth]{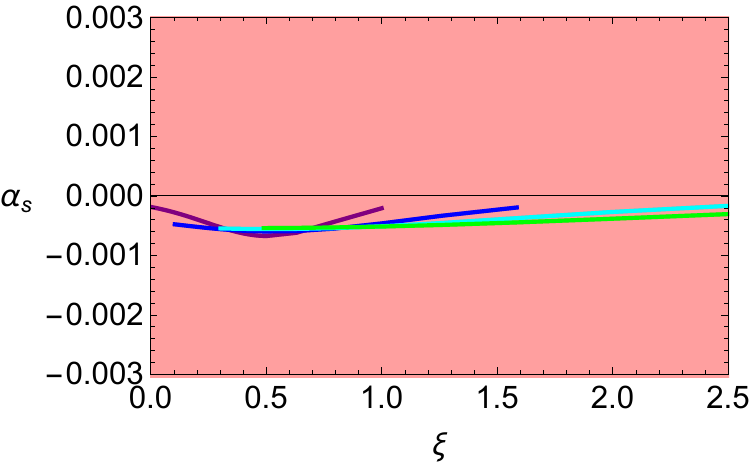}}   
    \caption{\em 
    $\alpha_\text{s}$ vs.~$n_\text{s}$ (a),  $\alpha_\text{s}$ vs.~$\xi$ (b) for $N_e=60$ and $r \lesssim 0.0383$ with 
    the same $\delta_f$ values and color code as in Fig.~\ref{fig:results:Palatini}.
     The pink areas represent the 1,2$\sigma$ allowed regions coming  from the Planck legacy data~\cite{Planck:2018vyg}, while the gray areas are the same as in Fig.~\ref{fig:results:Palatini:minus}.} 
    \label{fig:runningtilt:Palatini}%
\end{figure}

\begin{figure}[t!]
    \centering
     \includegraphics[width=0.49\textwidth]{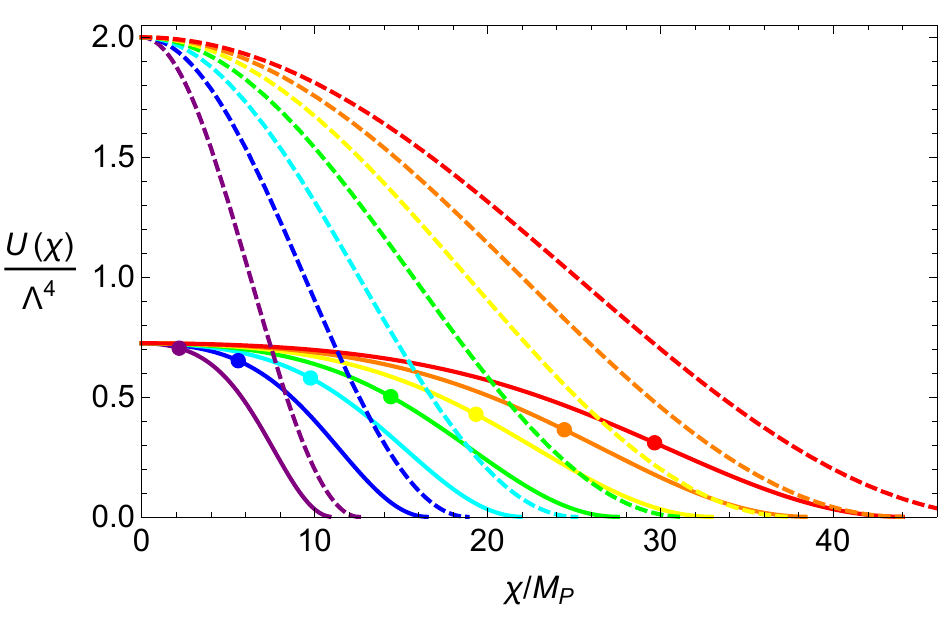}     
    \caption{\em $U(\chi)$ with $\xi \simeq 0.33$, $\beta_0=\tilde\xi=0$ (continuous) and $U(\chi)$ with $\xi=\beta_0=\tilde\xi=0$ (i.e the standard natural potential) (dashed)  for $\delta_f$ ranging from 4 (purple) to 16 (red) with steps of 2, displayed in rainbow colors. The bullets represents the corresponding points at $\chi_N$ with $N_e=60$.} 
    \label{fig:U:vs:chi:Palatini}%
\end{figure}

\subsubsection{$\xi > 0$ and $\beta(\phi)=0$} \label{subsec:Palatini}

In Fig.~\ref{fig:results:Palatini} we plot the corresponding results for $r$ and $n_\text{s}$ as functions of the parameters $\xi$ and $\delta_f$ when $N_e=60$  and $\delta_\Lambda$ is fixed so that the constraint \eqref{eq:As:exp} is satisfied.


Regardless of $\delta_f$, the predictions share some common properties. First, as expected, $r$ decreases by increasing $\xi$, while $n_\text{s}$ first reaches a maximum (whose numerical value is dependent on the actual value of $\delta_f$) and then decreases. A similar behaviour is also shown by $\delta_\Lambda$ as a function of $\xi$. Moreover,  $\delta_\Lambda$ also increases by increasing $\delta_f$.  At selected $\delta_f$ values, by increasing $\xi$ the predictions manage to reach the $2\sigma$ allowed region. In particular for $\delta_f=6$, they touch the border of the $1\sigma$ region. This happens for $\xi$ values around 1 or smaller. For $\delta_f \gtrsim 12$, no predictions enter the allowed region.

The results can be better understood by giving a look at the Einstein frame potential $U(\chi)$, which is plotted in Fig.~\ref{fig:U:vs:chi:Palatini} for selected benchmark points. 
For clarity the lines have been truncated at half period i.e.~$\chi=\chi(\phi_2)$. First of all we can see that the potential is lowered and flattened as an effect of the non-minimal coupling. As anticipated in Sec.~\ref{sec:potential}, the condition in~(\ref{eftC}) to rely on an effective-field-theory treatment of gravity is amply satisfied (see Fig.~\ref{fig:U:vs:chi:Palatini}). The main reason why this happens is the smallness of $\Lambda$ in Planck units, $\delta_\Lambda\ll 1$, (shown in Fig.~\ref{fig:results:Palatini}d); in the construction of Appendix~\ref{micro} such smallness is natural because it is generated by the smallness of quark-like masses (the only source of a chiral symmetry breaking) compared to a QCD-like confining scale.  The flattening of the potential usually comes with a lowering of $r$, which is confirmed by the predictions in Fig.~\ref{fig:results:Palatini}. We also see that by increasing $\delta_f$ (at fixed $\xi$), $\chi_N$ moves further and further away from the maximum of the potential towards steeper regions, implying an increase on $r$, which is confirmed by Fig.~\ref{fig:results:Palatini}b.

Since a relevant region of the parameters space falls within the 2$\sigma$ boundary of $r$ vs.~$n_\text{s}$, we study whether the constraints on the running of the spectral index $\alpha_\text{s}$ are satisfied. In Fig.~\ref{fig:runningtilt:Palatini} we plot the corresponding results as functions of the parameters $\xi$ and $\delta_f$. We considered only the parameters space that predicts $r \lesssim 0.0383$, which is the highest $r$-value in the 2$\sigma$ boundary line. The results for $\delta_f = 12, 14, 16$ are absent because they do not exhibit any region compatible with the $r$ vs.~$n_\text{s}$ constraints. This allows us to appreciate the different lines corresponding to the other values of $\delta_f$. We can see that in this case the stronger constraint is the one coming from $r$ vs.~$n_\text{s}$ and that all the points in agreement with the $r$ vs.~$n_\text{s}$ constraint are also in agreement with the $\alpha_\text{s}$ vs.~$n_\text{s}$ one. We also notice that in the allowed region, $\alpha_\text{s}$ is always predicted to be negative.

Overall we can conclude that the presence of the non-minimal coupling $\xi$ is beneficial to restore compatibility with data of natural inflation. This is also the case in the metric theory of natural inflation in the presence of the non-minimal coupling~\cite{Salvio:2023cry}, where $k(\phi)$ is given by~(\ref{eq:K(phi)2}) rather than~(\ref{eq:K(phi)b0}).

\subsection{$\xi = 0$ and $\tilde\xi > 0$}

\subsubsection{$\beta_0 \geq 0$}

\begin{figure}[t!]
     \subfloat[]{\includegraphics[width=0.45\textwidth]{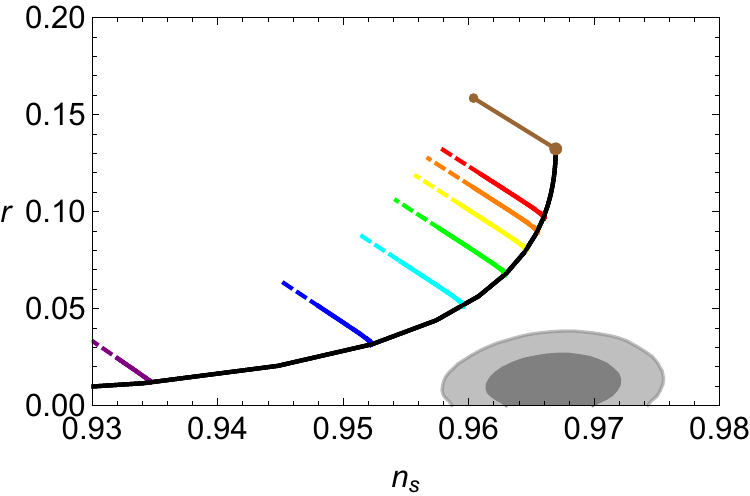}}%
    \subfloat[]{\includegraphics[width=0.45\textwidth]{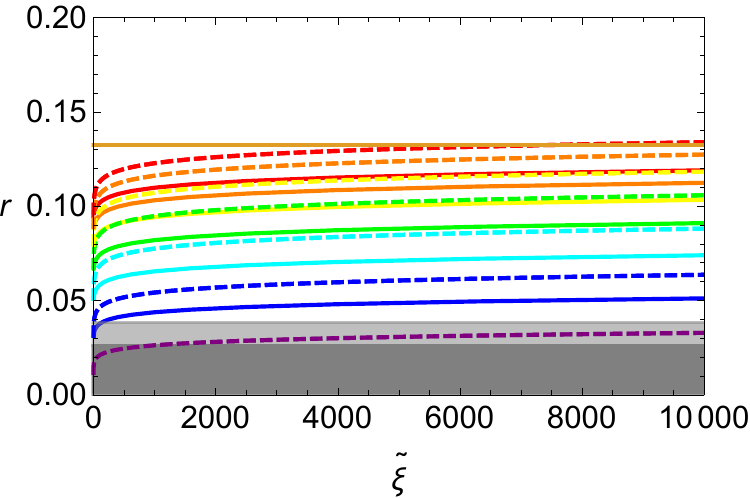}}%

    \subfloat[]{\includegraphics[width=0.45\textwidth]{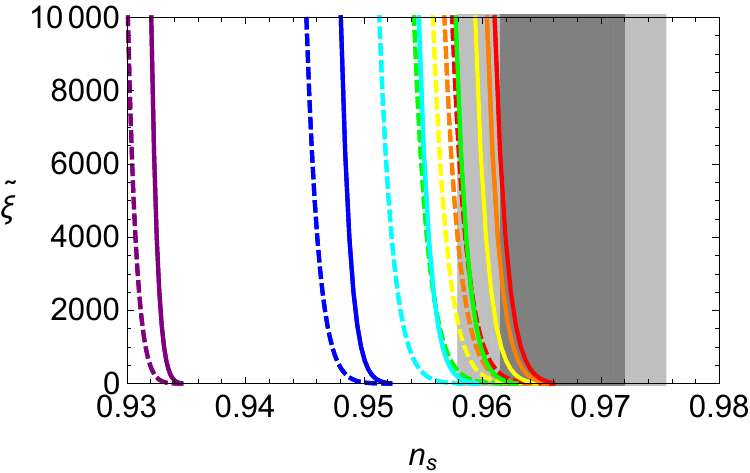}}%
    \subfloat[]{\includegraphics[width=0.45\textwidth]{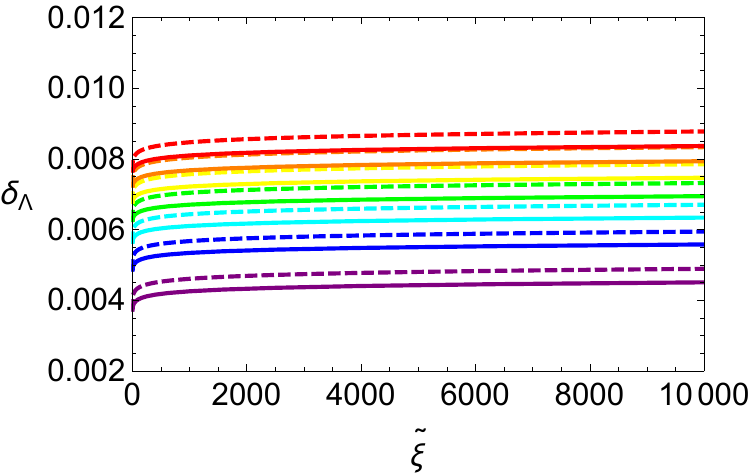}}%

    \caption{\em $r$ vs.~$n_\text{s}$ (a),  $r$ vs.~$\tilde\xi$ (b),  $\tilde\xi$ vs.~$n_\text{s}$ (c),  $\delta_\Lambda$ vs.~$\tilde\xi$ (d) for $N_e=60$ with $\xi=0$, $\beta_0=0$ (dashed) and  $\beta_0=2 M_P^2$ (continuous) for $\delta_f$ ranging from 4 (purple) to 10 (red) with steps of 1, displayed in rainbow colors. The gray, brown and black color codes are the same as in Fig.~\ref{fig:results:Palatini:minus}.}
    %
    \label{fig:results:pos:xi:prime}
\end{figure}

In Fig.~\ref{fig:results:pos:xi:prime} we plot the corresponding results for $r$ and $n_\text{s}$ versus the parameters $\tilde\xi$ and $\delta_f$ when  $\beta_0 \geq 0$, $N_e=60$  and $\delta_\Lambda$ is fixed so that the constraint \eqref{eq:As:exp} is satisfied.


We can see that the $\beta_0>0$ and $\beta_0=0$ results overlap with each other. Therefore, the effect of having $\beta_0>0$ is just to shift the results at higher $\tilde\xi$ values. Unfortunately, with $\tilde\xi$ increasing in both cases $r$ ($n_\text{s}$) increases (decreases) leading the prediction even more away from the allowed region.  Because of this, we truncated the study at $\tilde\xi=10000$ and concluded that the configuration with $\xi = 0$ and $\beta_0 \geq 0$ is excluded by data 
and the computation of the predictions for $\alpha_\text{s}$ is not needed.

\subsubsection{$\beta_0 < 0$} \label{subsec:minus:xi:prime:2}

\begin{figure}[p!]
     \subfloat[]{\includegraphics[width=0.45\textwidth]{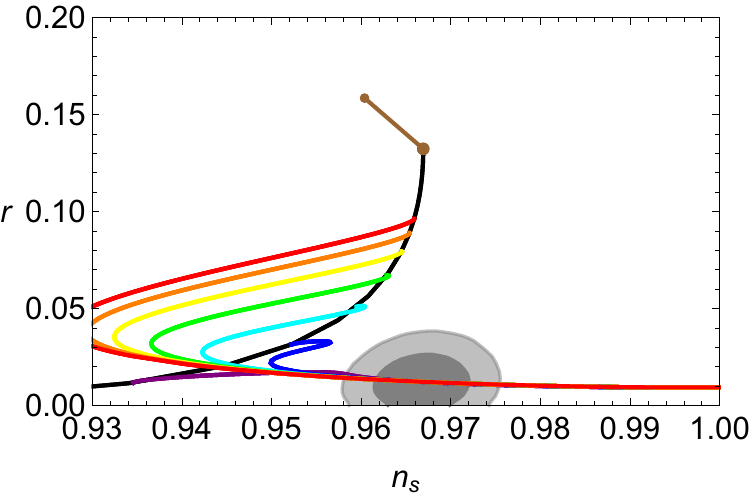}}%
    \subfloat[]{\includegraphics[width=0.45\textwidth]{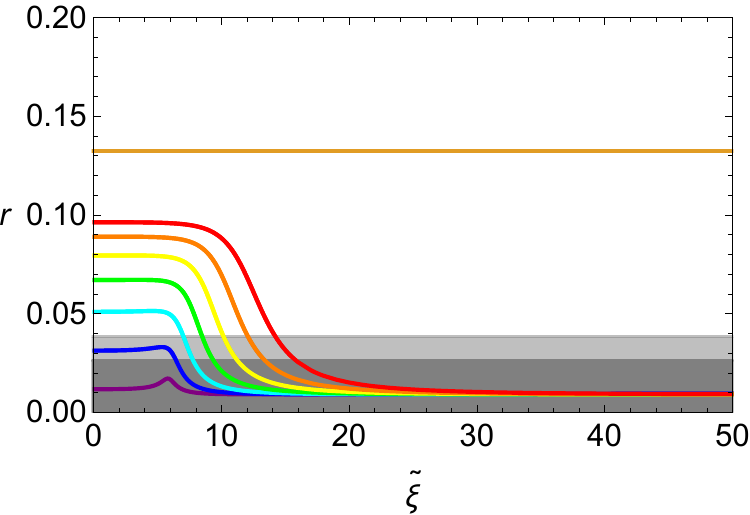}}%

   \hspace*{0.1cm} \subfloat[]{\includegraphics[width=0.44\textwidth]{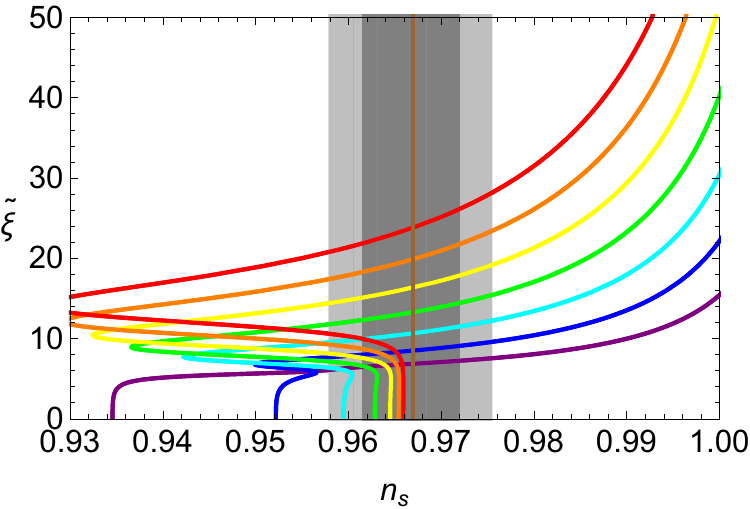}}%
    \subfloat[]{\includegraphics[width=0.45\textwidth]{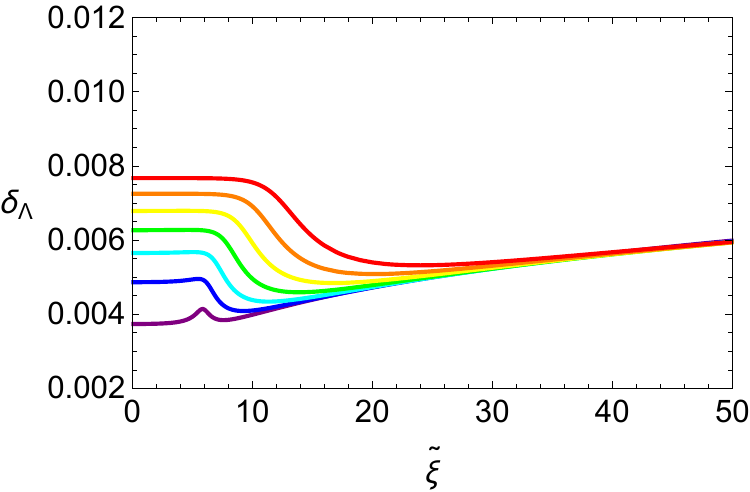}}%
    \caption{\em $r$ vs.~$n_\text{s}$ (a),  $r$ vs.~$\tilde\xi$ (b),  $\tilde\xi$ vs.~$n_\text{s}$ (c),  $\delta_\Lambda$ vs.~$\tilde\xi$ (d) for $N_e=60$ with $\xi=0$, $\beta_0=-6 M_P^2$ for $\delta_f$ ranging from 4 (purple) to 10 (red) with steps of 1, displayed in rainbow colors. The gray, brown and black color codes are the same as in Fig.~\ref{fig:results:Palatini:minus}.}
    %
    \label{fig:results:minus:xi:prime:2}%
\end{figure}

\begin{figure}[pb!]
     \subfloat[]{\includegraphics[width=0.5\textwidth]{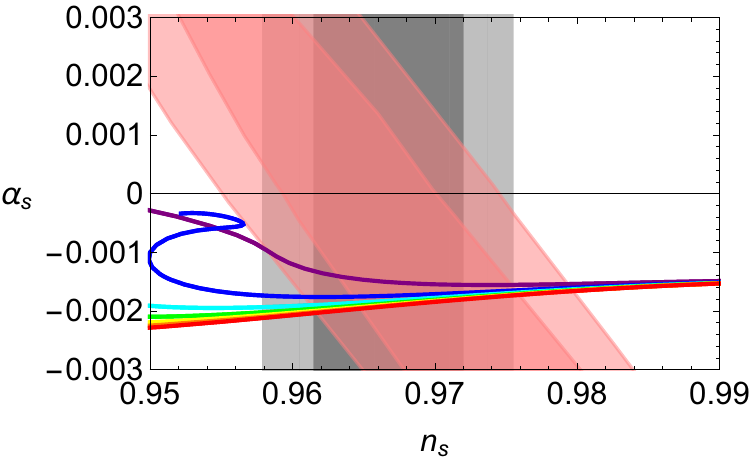}}%
    \subfloat[]{\includegraphics[width=0.47\textwidth]{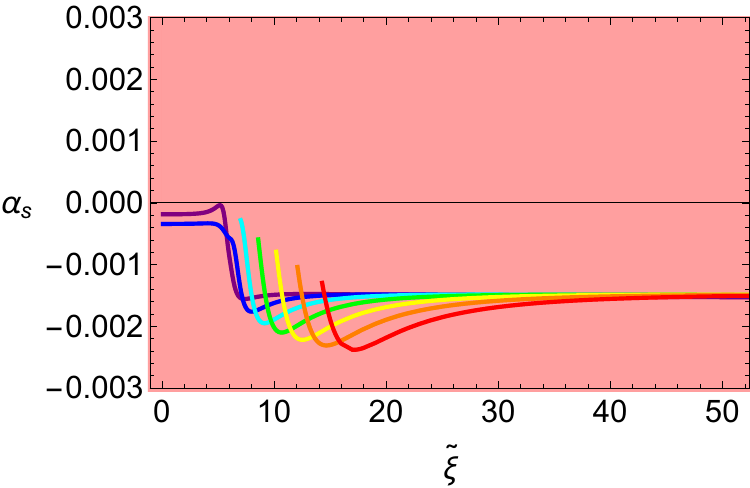}}   
    \caption{\em 
    $\alpha_\text{s}$ vs.~$n_\text{s}$ (a),  $\alpha_\text{s}$ vs.~$\tilde\xi$ (b) for $N_e=60$ and $r \lesssim 0.0383$ with $\xi=0$, $\beta_0=-6 M_P^2$ for 
    the same $\delta_f$ values and color code as in Fig.~\ref{fig:results:minus:xi:prime:2}.
      The pink and gray areas are the same as in Fig.~\ref{fig:runningtilt:Palatini}.} 
    \label{fig:runningtilt:minus:xi:prime:2}%
\end{figure}
In Fig.~\ref{fig:results:minus:xi:prime:2} we plot the corresponding results for $r$ and $n_\text{s}$ versus the parameters $\tilde\xi$ and $\delta_f$ when $\beta_0=-6 M_P^2$, $N_e=60$  and $\delta_\Lambda$ is fixed so that the constraint \eqref{eq:As:exp} is satisfied.

\begin{figure}[t!]
    \centering
    \subfloat[]{   \includegraphics[width=0.45\textwidth]{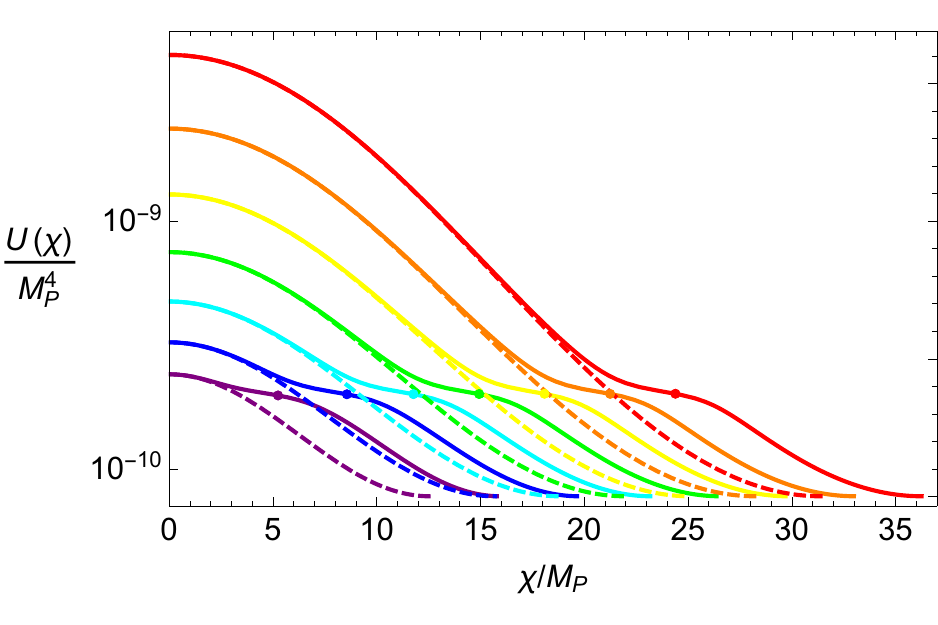}}
    \quad 
    \subfloat[]{   \includegraphics[width=0.45\textwidth]{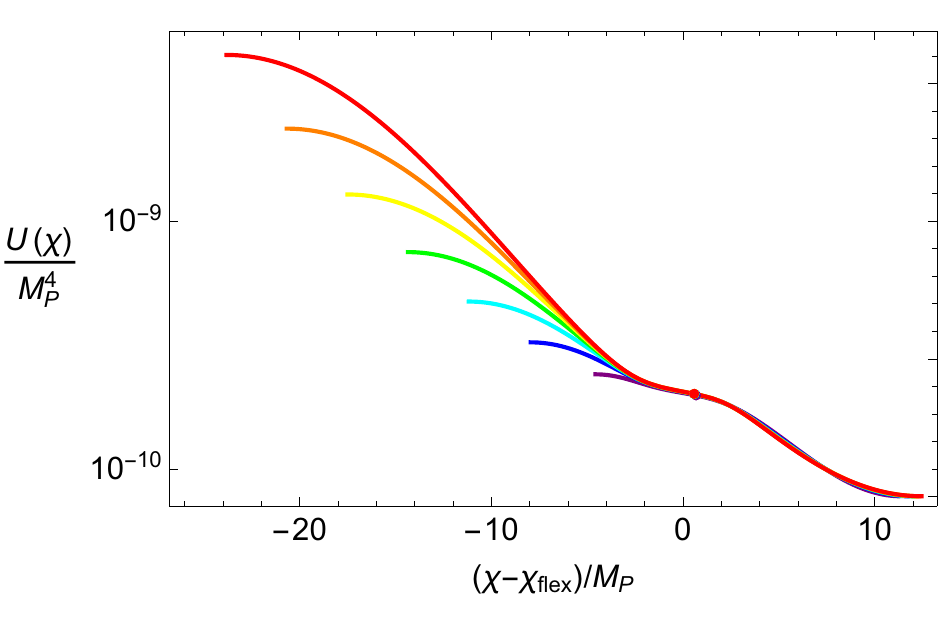}}
    \caption{\em $U(\chi)$ (a) and $U(\chi - \chi_\text{flex})$ (b) with $\xi=0$, $\beta_0=-6 M_P^2$  and $n_\text{s} \simeq 0.97$ (continuous) and $U(\chi)$ with $\xi=\beta_0=\tilde\xi=0$ (dashed)  for $\delta_f$ ranging from 4 (purple) to 10 (red) with steps of 1, displayed in rainbow colors. In the continuous lines $\tilde\xi$ varies with $f$ such that $n_\text{s}\simeq 0.97$. The bullets represents the corresponding points at $\chi_N$ with $N_e=60$.}
    \label{fig:U:vs:chi:MAG}%
\end{figure}


We can see that by increasing $\tilde\xi$, $r$ decreases in most of the chosen configuration but $\delta_f=4,5$ where, instead, $r$ vs.~$\tilde\xi$ exhibits first an increasing behaviour and then a decreasing one like in all the other $\delta_f$'s. On the other hand, the behaviour of $n_\text{s}$ vs.~$\tilde\xi$  is more oscillating (showing first a short increasing phase, then a decreasing one and finally a new increasing phase) for all the chosen configurations but for $\delta_f=4$ where $n_\text{s}$ appears to just increase with $\tilde\xi$. At big enough $\tilde\xi$, the results in the region $r \lesssim 0.015$ and $n_\text{s} \gtrsim 0.958$  overlap, but they are still dependent on $f$. This is visible in Fig.~\ref{fig:results:minus:xi:prime:2}(c). This means that by changing $f$, we can find a $\tilde\xi$ so that the results instead do not change. 

 Once again, it is worth studying the predictions for $\alpha_\text{s}$. In Fig.~\ref{fig:runningtilt:minus:xi:prime:2} we plot the corresponding results as functions of the parameters $\tilde\xi$ and $\delta_f$. The same cut in $r$ as in Fig.~\ref{fig:runningtilt:Palatini} has been implemented. We can see that the constraint on $\alpha_\text{s}$ reduces the allowed parameter space.
Again all the predicted values for $\alpha_\text{s}$ are negative but now larger in absolute value than those of the $\beta=0$ case in Fig.~\ref{fig:runningtilt:Palatini}.

To get a better understanding of the large $\tilde\xi$ behaviour, 
 we plot for selected benchmark points $U(\chi)$ centered at the origin and centered around the corresponding inflection point respectively in Fig.~\ref{fig:U:vs:chi:MAG}(a) and 
in Fig.~\ref{fig:U:vs:chi:MAG}(b).
For clarity the lines have been truncated at half period i.e.~$\chi = \chi(\phi_2)$. For the selected points, $\chi_N$ is close to the inflection point of the potential. Moreover, it is hard to distinguish the different potentials around (and after) the inflection points. The equation for such a points is
\be
 U''(\chi_\text{flex})=0 \label{eq:flex:point}
\ee
In terms of $\phi$ \eqref{eq:flex:point} can be rewritten as
\be 
 \frac{1}{2} \frac{k'(\phi_\text{flex})}{k(\phi_\text{flex})} = \frac{U''(\phi_\text{flex})}{U'(\phi_\text{flex})} \, . \label{eq:flex:point:2}
\ee
Using Eq.~\eqref{eq:U:phi} and imposing $\xi=0$, \eqref{eq:flex:point:2} becomes
 \be 
 \frac{1}{2} \frac{k'(\phi_\text{flex})}{k(\phi_\text{flex})} = \frac{1}{f} \cot\left( \frac{\phi_\text{flex}}{f} \right) \, . \label{phiflexeq}
\ee
The right-hand side of this equation turns out to be strongly suppressed, such that we can very well approximate \eqref{eq:flex:point:2} as $k'(\phi) \simeq 0$. As a result, the inflection point of $U$ and the maximum point of $k$ are numerically almost indistinguishable. 
As we can see in Fig.~\ref{fig:U:vs:chi:MAG}(b) a change in $f$ can be compensated by a change in $\tilde\xi$ so that the shape of  $U$ around $\phi_\text{peak}$ stays unchanged (the constraint between $f$ and $\tilde\xi$ can be formalized as $U(\phi_\text{peak})=$ constant). 
This is reflected in the inflationary results as well, where all the lines overlap when $\phi_N$ is around $\phi_\text{peak}$.

As anticipated in Sec.~\ref{sec:potential}, once again we note that the condition in~(\ref{eftC}) to rely on an effective-field-theory treatment of gravity is amply satisfied, as shown by Fig.~\ref{fig:U:vs:chi:MAG}. Also here the main reason why this happens is the smallness of  $\delta_\Lambda$, (shown here in Fig.~\ref{fig:results:minus:xi:prime:2}d), which is natural in the construction of Appendix~\ref{micro}.

 
Overall we can conclude that the presence of the non-minimal coupling $\tilde\xi$ combined with a negative $\beta_0$ can restore the compatibility with data of natural inflation.

\begin{figure}[p!]
     \subfloat[]{\includegraphics[width=0.45\textwidth]{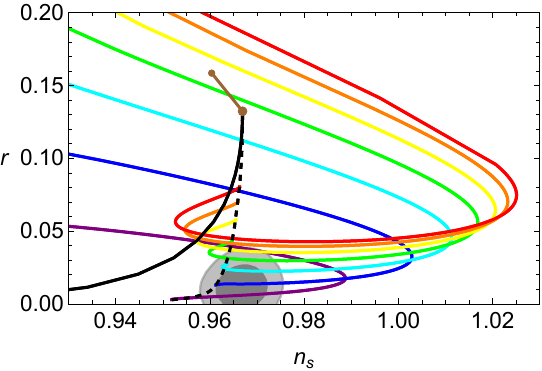}}%
   ~~ \subfloat[]{\includegraphics[width=0.44\textwidth]{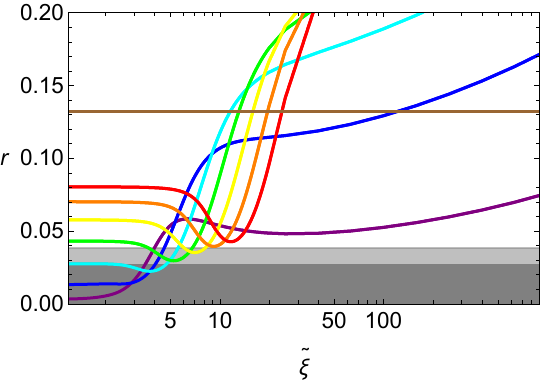}}%

    \subfloat[]{\includegraphics[width=0.45\textwidth]{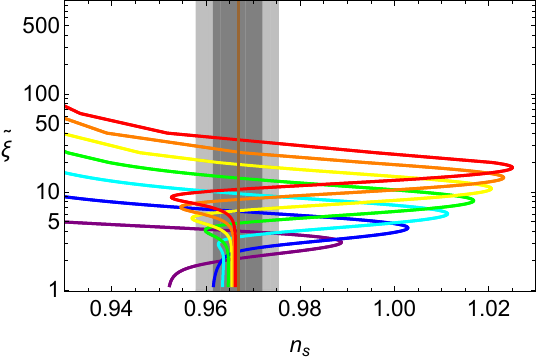}}%
  ~~  \subfloat[]{\includegraphics[width=0.45\textwidth]{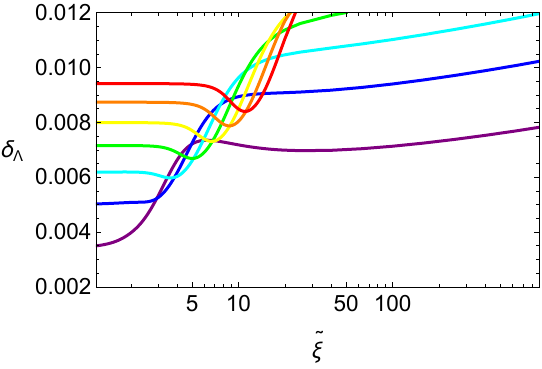}}%
    \caption{\em $r$ vs.~$n_\text{s}$ (a),  $r$ vs.~$\tilde\xi$ (b),  $\tilde\xi$ vs.~$n_\text{s}$ (c),  $\delta_\Lambda$ vs.~$\tilde\xi$ (d) for $N_e=60$ with $\xi=\frac{1}{3}$, $\beta_0=-2 M_P^2$ for $\delta_f$ ranging from 3 (purple) to 15 (red) with steps of 2, displayed in rainbow colors. The gray, brown and black color codes are the same as in Fig.~\ref{fig:results:Palatini:minus}.}
    %
    \label{fig:results:Palatini:minus:xi:prime}%
\end{figure}

\begin{figure}[pb!]
     \subfloat[]{\includegraphics[width=0.47\textwidth]{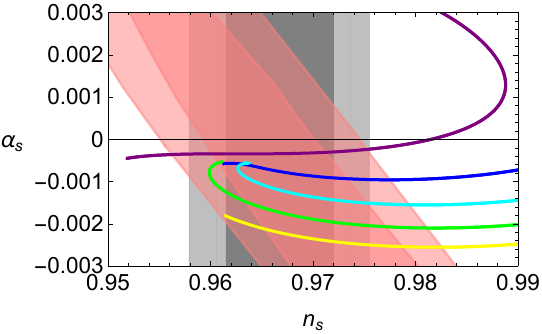}}%
    \subfloat[]{\includegraphics[width=0.45\textwidth]{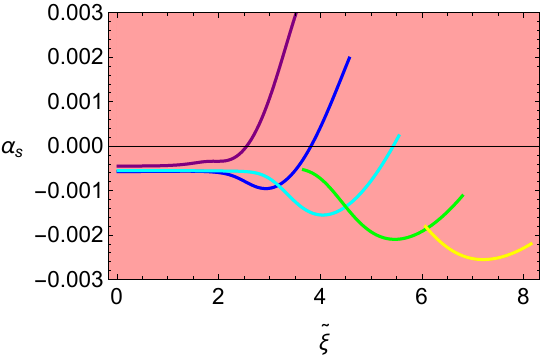}}   
    \caption{\em $\alpha_\text{s}$ vs.~$n_\text{s}$ (a),  $\alpha_\text{s}$ vs.~$\tilde\xi$ (b) for $N_e=60$ and $r \lesssim 0.0383$ with $\xi=\frac{1}{3}$, $\beta_0=-2 M_P^2$ for 
    the same $\delta_f$ values and color code of Fig.~\ref{fig:results:Palatini:minus:xi:prime}.
     The pink and gray areas are the same as in Fig.~\ref{fig:runningtilt:Palatini}.} 
    \label{fig:runningtilt:Palatini:minus:xi:prime}%
\end{figure}

\subsection{$\xi >0$ and $\tilde\xi > 0$}

\begin{figure}[p!]
     \subfloat[]{\includegraphics[width=0.45\textwidth]{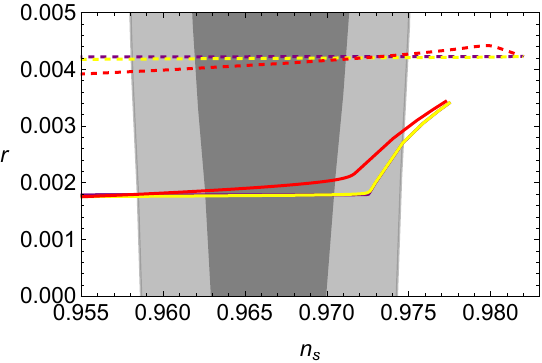}}%
   ~~ \subfloat[]{\includegraphics[width=0.44\textwidth]{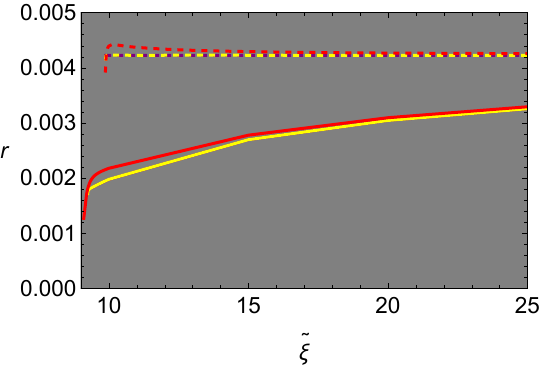}}%

   \hspace{0.6cm} \subfloat[]{\includegraphics[width=0.45\textwidth]{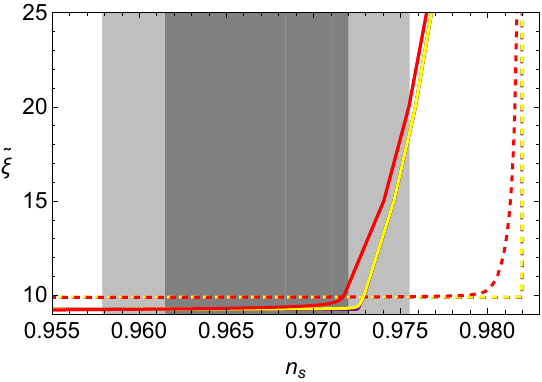}}%
  ~~  \subfloat[]{\includegraphics[width=0.45\textwidth]{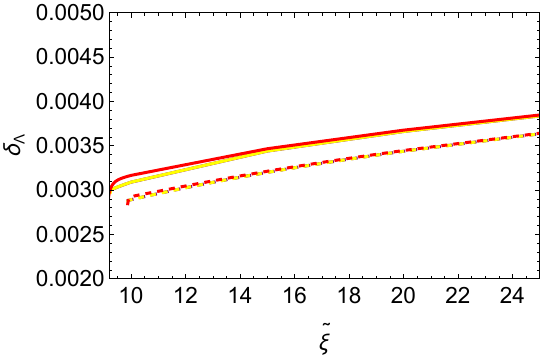}}%

    \caption{\em $r$ vs.~$n_\text{s}$ (a),  $r$ vs.~$\tilde\xi$ (b),  $\tilde\xi$ vs.~$n_\text{s}$ (c),  $\delta_\Lambda$ vs.~$\tilde\xi$ (d) for $N_e=60$ with $\beta_0=-10 M_P^2$ for $\delta_f=10^{-2}$ (purple), $\delta_f= 10^{-1}$ (yellow) and $\delta_f= 1$ (red) when $\xi=\frac{1}{3}$ (continuous) or $\xi=0$ (dashed). The gray color codes are the same as in Fig.~\ref{fig:results:Palatini:minus}.}
    %
    \label{fig:results:Palatini:minus:xi:prime:low:f}%
\end{figure}

\begin{figure}[pb!]
     \subfloat[]{\includegraphics[width=0.47\textwidth]{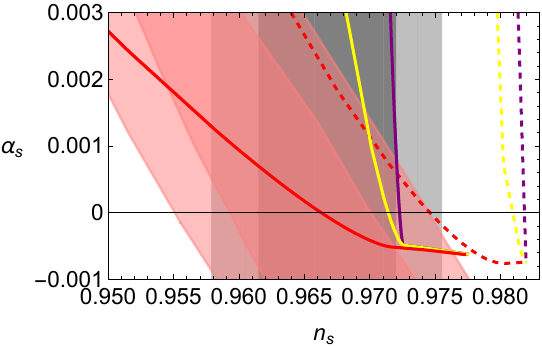}}%
   \quad
    \subfloat[]{\includegraphics[width=0.47\textwidth]{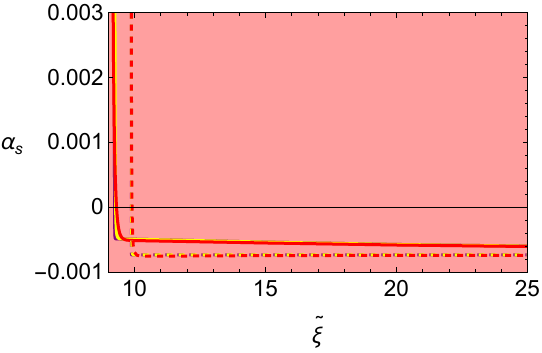}}   
    \caption{\em $\alpha_\text{s}$ vs.~$n_\text{s}$ (a),  $\alpha_\text{s}$ vs.~$\tilde\xi$ (b) for $N_e=60$ with $\beta_0=-10 M_P^2$ for the same values and color codes of $\delta_f$ and $\xi$ as in Fig.~\ref{fig:results:Palatini:minus:xi:prime:low:f}.
    The pink and gray areas are the same as in Fig.~\ref{fig:runningtilt:Palatini}.} 
    \label{fig:runningtilt:Palatini:minus:xi:prime:low:f}%
\end{figure}

In  Figs.~\ref{fig:results:Palatini:minus:xi:prime} and ~\ref{fig:runningtilt:Palatini:minus:xi:prime} we plot the corresponding results for $r$, $n_\text{s}$ and $\alpha_\text{s}$ versus the parameters $\tilde\xi$ and $\delta_f$ when $\xi=\frac{1}{3}$, $\beta_0=-2 M_P^2$, $N_e=60$  and $\delta_\Lambda$ is fixed so that the constraint \eqref{eq:As:exp} is satisfied.


The results are clearly a combination of the behaviour pattern shown 
in subsections \ref{subsec:Palatini} and \ref{subsec:minus:xi:prime:2}. The most interesting result is the appearance of regions where a positive $\alpha_\text{s}$ is predicted. However, these regions are ruled out because of the constraint on  $r$ vs.~$n_\text{s}$, but for $\delta_f=3$ where the positive $\alpha_\text{s}$ is excluded by the $\alpha_\text{s}$ vs.~$n_\text{s}$ constraint. 

Overall, as expected, compatibility with data is achieved quite easily. 

 Moreover, this case is  interesting also because it allows us to satisfy the experimental constraints even when $f$ is sub-Planckian. The corresponding results are shown in Figs.~\ref{fig:results:Palatini:minus:xi:prime:low:f} and \ref{fig:runningtilt:Palatini:minus:xi:prime:low:f}. For numerical convenience the study considers only values of $\tilde\xi$ between approximately 10 and 25. The lines representing to the results corresponding to $\delta_f=10^{-2}$ and $\delta_f=10^{-1}$ are essentially undistinguishable apart for the $\alpha_s$ vs. $n_s$ plot. We can see that when $\xi=0$, with a large enough $\tilde \xi$, the $r$ vs.  $n_s$ predictions at very low $\delta_f$ are again compatible with data. However, in this case $\alpha_s$ is very large and essentially always out from the allowed region for sub-Planckian $f$. This was somehow expected because in standard natural inflation the second slow-roll parameter $\eta$ is very large in absolute value at low $\delta_f$ (potentially even breaking the slow-roll approximation), implying a very small $n_s$. In order to achieve compatibility with data, $n_s$ is varying very fast causing its running $\alpha_s$ to be out of the allowed region. The issue can be solved by using $\xi > 0$, which restores compatibility with data by moving the large $\alpha_s$ region towards smaller $n_s$ values.

\section{Conclusions}\label{Conclusions}

In this paper we have constructed and analysed a metric-affine realization of natural inflation.

  We have focused on the low-energy (two-derivative) metric-affine inflationary theory containing only the graviton and the PNGB inflaton in the particle content. In particular, the connection  does not carry additional degrees of freedom. This theory, whose action is given in\footnote{It is worth mentioning that~(\ref{SNIstart}) is also the low-energy metric-affine inflationary theory for any single-field scenario with generic potential $V$, which is not necessarily the natural-inflation potential in~(\ref{eq:V(phi)}).}~(\ref{SNIstart}), features, besides the PNGB potential,  non-minimal couplings between the PNGB and the two linear-in-curvature invariants that one can construct in metric-affine gravity: ${\cal R}$ and $\tilde{\cal R}$, defined in~(\ref{RRpdef}). We have integrated out the distorsion to find an explicit metric theory, with action given in~(\ref{SNIE}), where the connection is given by the LC formula.  Moreover, in Appendix~\ref{micro} we have found an explicit UV completion for the cosine potential and both non-minimal couplings in~(\ref{eq:V(phi)}) and~(\ref{alphabetaUV}), respectively: by UV completion we mean that these functions emerge from the microscopic dynamics of a quantum field theory that is renormalizable on Minkowski spacetime. The UV completion of Einstein gravity is beyond the scope of the present paper; perhaps it can be achieved 
  in string theory or
 in asymptotically free/safe theories
 featuring higher-derivative terms. 

  We have performed an analytical study of the potential of the canonically normalized inflaton, $\chi$, which is defined in terms of the original PNGB $\phi$ through the kinetic function $k$, see Eqs.~(\ref{eq:K(phi)}) and~(\ref{eq:define.Chi}).  We have analytically shown  that $k$ develops a high peak for large and opposite values of $\beta_0$ and $\tilde\xi$, defining the non-minimal coupling between the PNGB and the Holst invariant, the second formula in~(\ref{alphabetaUV}). This corresponds to a plateau in the potential of $\chi$, which has been found explicitly.
  
  Finally, we have performed an analysis of the inflationary predictions of this theory for the potential and non-minimal couplings in~(\ref{eq:V(phi)}) and~(\ref{alphabetaUV}), respectively, because, as shown, they admit an explicit UV completion.  For these models,  we have found regions of the parameter space where the inflationary predictions agree with the most recent observations performed by the Planck, BICEP and Keck collaborations at the $2\sigma$ level. We have found that in order to enter the $1\sigma$ region it is necessary (and sufficient) to have a finite value of the Barbero-Immirzi parameter (a sizable $|\beta_0|$) and a sizable $\tilde\xi$ (with sign opposite to the Barbero-Immirzi parameter). This is precisely the region of parameter space for which the potential of $\chi$ develops a plateau. It is indeed the presence of this plateau that renders the theory observationally viable. As far as $\delta_f$ is concerned, the results of this work are obtained considering  the window $\delta_f\in[10^{-2},16]$. We have shown that this scenario can be compatible with observations  for both trans-Planckian and sub-Planckian values of $f$.

\subsubsection*{Acknowledgments}
A.R.~thanks I.D. Gialamas for useful discussions. The work of A.R.~was supported by the Estonian Research Council grants PRG1055,  RVTT3, RVTT7 and the CoE program TK202 ``Fundamental Universe". The work of A.S.~was also supported by the grant DyConn from the University of Rome Tor Vergata. This article is based upon work from COST Actions COSMIC WISPers CA21106 and CosmoVerse CA21136, supported by COST (European Cooperation in Science and Technology).

 \appendix
  \section{A microscopic origin of natural metric-affine inflation}\label{micro}

  Let us present a possible microscopic origin not only of $V$, which has been briefly discussed in e.g.~\cite{Freese:1990rb}, but also of $\alpha$ and $\beta$. A similar discussion but for the metric theory of natural inflation with non-minimal coupling in~(\ref{metricNI}) has been made in~\cite{Salvio:2021lka}.
  
  The simplest possibility, which we treat here, is to consider a version of QCD with a confinement scale $f$ around the Planck scale and with three flavors of  tilde-quarks: $\tilde q=\{\tilde u, \tilde d, \tilde s\}\equiv \{\tilde q_1, \tilde q_2, \tilde q_3\}$, where the tilde distinguishes from the analogous QCD quantities.
 Like in ordinary QCD the strong dynamics forms condensates with a scale  $\tau$~\cite{Weinberg2}, 
\be \langle  \bar{\tilde q}'_i \tilde q'_j \rangle = -\tau \delta_{ij}, \qquad \langle\bar{\tilde q}'_i \gamma_5\tilde q'_j \rangle =0, \label{qVEV} \ee  
where $\langle\cdot\rangle$ represents the vacuum expectation value and $\tilde q'_i$  are the Goldstone-free quark fields:
\be \tilde q' = \exp(i \gamma_5 B/(\sqrt{2}f)) \tilde q. \label{qp}\ee 
 Also, $f$ is analogous to the pion decay constant and  $B$ is the Hermitian matrix containing the tilde-mesons (with canonically normalized kinetic terms)
\be \label{Bf}   B\equiv \left(\baccc \frac{\tilde\pi^{0}}{\sqrt{2}} +\frac{\tilde\eta^0}{\sqrt{6}}& \tilde\pi^+ & \tilde K^+ \\
(\tilde\pi^+)^\dagger & -\frac{\tilde\pi^{0}}{\sqrt{2}} +\frac{\tilde\eta^0}{\sqrt{6}} & \tilde K^0 \\
(\tilde K^+)^\dagger & (\tilde K^0)^\dagger & -\sqrt{\frac{2}{3}}\tilde\eta^0 \ea \right). 
\ee
These scalars are the Goldstone bosons associated with the breaking of the axial part of the global ${\rm SU(3)}_{\rm f}$  flavor group (acting on $\{\tilde u, \tilde d, \tilde s\}$). Just like in QCD, one can add quark mass terms that explicitly break the axial part of ${\rm SU(3)}_{\rm f}$  flavor group:
\be \mathscr{L}_{\rm mass} = \bar{\tilde q} M_q\tilde q = \bar{\tilde q}'  \exp(-i \gamma_5 B/(\sqrt{2}f)) M_q \exp(-i \gamma_5 B/(\sqrt{2}f)) \tilde q', \ee
where $M_q$ is the tilde-quark mass matrix. The group ${\rm SU(3)}_{\rm f}$ is an approximate symmetry when the elements of $M_q$ are small compared to $f$.
The tilde-meson potential can be computed  from $ \mathscr{L}_{\rm mass}$ using~(\ref{qVEV}). 
In general this potential turns out to be
\be V = \tau\,  {\rm Tr}\left[ \cos\left(\sqrt{2}B/f\right) M_q\right] +\Lambda_0,\label{VNq}\ee
where $\Lambda_0$ is a  real constant.

Using standard effective field theory methods~\cite{Weinberg2} and taking $M_q$ diagonal for simplicity,
\be M_q = {\rm diag}(m_{\tilde u},m_{\tilde d},m_{\tilde s}), \ee
one finds the following spectrum of the PNGBs
\bea  m^2_{\tilde K^0} &=& \frac{\tau}{f^2}(m_{\tilde d} + m_{\tilde s}), \label{K0} \\ 
m^2_{\tilde K^+} &=& \frac{\tau}{f^2}(m_{\tilde u} + m_{\tilde s}), \label{Kp} \\ 
m^2_{\tilde \pi^0} &=& m^2_{\tilde \pi^+}  = \frac{\tau}{f^2}(m_{\tilde u} + m_{\tilde d}),  \label{pi0}\\
m^2_{\tilde \eta^0} &=& \frac{\tau}{f^2}\left(\frac{m_{\tilde u}+m_{\tilde d} + 4 m_{\tilde s}}{3}\right). \label{eta0}
\eea
By using the known values of the meson masses, the up and down quark masses and the pion decay constant one obtains
\be \tau \sim 30 f^3 \label{kappaf}. \ee
This relation should also be approximately true in this variant of QCD 
as long as the elements of $M_q$ are  much smaller than $f$.

 Now, by choosing  $m_{\tilde  u} \gg m_{\tilde d}, m_{\tilde  s}$ we obtain that the lightest pseudo-Goldstone boson is the complex scalar $\tilde K^0$.
During inflation  we can parameterize it as 
\be\tilde K^0 = \frac{\phi}{\sqrt{2}} \exp(i\omega_\phi/f),\ee where $\phi$ is the real inflaton field we have introduced in~(\ref{SNIstart}) and $\omega_\phi$ is some angular field. To compute the low energy potential for $\tilde K^0$ we can  set all other (heavy) tilde-meson fields to zero in~(\ref{Bf}):
 \be \label{Bf2}   B= \left(\baccc 0& 0 &0 \\
0 & 0 &\frac{\phi}{\sqrt{2}}  \,  e^{i\omega_\phi/f} \\
0 &\frac{\phi}{\sqrt{2}}  \, e^{-i\omega_\phi/f} & 0 \ea \right). 
\ee
In this case the eigenvalues of $B$ are $0$,  $\phi/\sqrt{2}$ and $-\phi/\sqrt{2}$ so
 \be \label{cBf}  \cos\left(\frac{\sqrt{2}B}{f}\right)= P +\cos\left(\frac{\phi}{f}\right) (1-P),
\ee
where $P =$ diag$(1,0,0)$.
From~(\ref{VNq}) the potential is
 \be V(\phi) = \tau (m_{\tilde d}+m_{\tilde s}) \cos\left(\frac{\phi}{f}\right) +\Lambda_0+\tau m_{\tilde u}.\ee
 Comparing this expression with~(\ref{eq:V(phi)}) we obtain  $\Lambda = [\tau (m_{\tilde d}+m_{\tilde s})]^{1/4}$  and  $\Lambda_0 = \Lambda^4+\Lambda_{\rm cc}-\tau m_{\tilde u}$. 
  
Sizable $\alpha$ and $\beta$ could appear, on the other hand, due to non-minimal interactions between $\tilde q$ and gravity predicted at low energies by some theories of quantum gravity. Consider e.g.~the effective low energy couplings 
\be \mathscr{L}_{q{\cal R}}  = \frac{m_P^2}{2} {\cal R} -\frac{1}{2\bp}\bar{\tilde q} J \tilde q {\cal R}-\frac{1}{2\bp}\bar{\tilde q} J' \tilde q \tilde{\cal R} = \frac{m_P^2}{2} {\cal R}-\frac{J_{ij}}{2\bp}\bar{\tilde q}_i \tilde q_j {\cal R}-\frac{J'_{ij}}{2\bp}\bar{\tilde q}_i \tilde q_j \tilde{\cal R}, \label{LqRs}\ee
where $m_P^2 {\cal R}/2$ is a ``bare" metric-affine Einstein-Hilbert term and $J$ and $J'$ are $3\times 3$ matrices of constant real coefficients. By using~(\ref{qVEV}) and~(\ref{qp}) one finds the following effective term proportional to ${\cal R}$
\be \frac{m_P^2}{2} {\cal R}+ \frac{\tau}{2\bp} {\rm Tr}\left[\cos\left(\sqrt{2}B/f\right) J\right] {\cal R}+ \frac{\tau}{2\bp} {\rm Tr}\left[\cos\left(\sqrt{2}B/f\right) J'\right] \tilde{\cal R},\ee
Now, using~(\ref{cBf}), one obtains~(\ref{alphabetaUV}) 
having identified
\be   \xi   = \frac{\tau}{M_P^3}\Tr(J-PJ), \qquad m_P^2 =   (1+\xi)\bp^2-\frac{\tau}{M_P}\Tr(PJ) \ee
and 
\be   \tilde\xi   = \frac{\tau}{M_P^3}\Tr(J'-PJ'), \qquad \beta_0   = \frac{\tau}{2M_P}\Tr(PJ') -\frac{M_P^2\tilde\xi}{2}. \ee

 \vspace{1cm}
\footnotesize
\begin{multicols}{2}

\end{multicols}

\end{document}